\documentclass[11pt]{article}
\usepackage{cite}
\usepackage{amsmath,amsfonts,amssymb}
\pdfoutput=1
\usepackage[small,bf,hang]{caption}
\usepackage{slashed}
\input epsf.sty
\usepackage{epsfig}
\usepackage[titletoc,toc]{appendix}
\usepackage{xcolor}
\usepackage{verbatim}
\usepackage{multirow}

\graphicspath{ {./images/} }

\def\hybrid{
        \topmargin -20pt
        \oddsidemargin 0pt
        \headheight 0pt \headsep 0pt
        \textwidth 6.55in 
        \textheight 9.5in 
        \marginparwidth .875in
        \parskip 5pt plus 1pt \jot = 1.5ex}

\hybrid

 \csname
@addtoreset\endcsname{equation}{section}

\newcommand{\M}{{\cal M}}

\def\moth{\mathsurround=0pt}
\newdimen\zo \zo=0pt

\def\tick{\leaders\hrule height 0.5ex depth 0pt \hskip 0.5pt}
\def\upboxfill{$\moth \setbox\zo\hbox{\tick}%
  \hskip 3pt\hbox to 0pt{$\tick$\hss}\hrulefill \hbox to 7.5pt{$\tick$\hss}$}

\def\dtick{\leaders\hrule height .34pt depth 0.5ex \hskip 0.5pt}
\def\downboxfill{$\moth \setbox\zo\hbox{\dtick}%
  \hskip 2pt\hbox to 0pt{$\dtick$\hss}\hrulefill \hbox to 2pt{$\dtick$\hss}$}

\def\Real{{\mathbb R}}
\def\Comp{{\mathbb C}}

\def\bec{\begin{center}}
\def\ec{\end{center}}

\def\f{\phi}

\def\vf{\varphi}

\def\S{\Sigma}

\def\be{\begin{equation}}
\def\ee{\end{equation}}
\def\bea{\begin{eqnarray}}
\def\eea{\end{eqnarray}}
\def\ba{\begin{array}}
\def\ea{\end{array}}

\def\V{\mathcal{V}}

\usepackage{hyperref}

\begin{document}

\begin{titlepage}

\rightline{\tt MIT-CTP-5494}
\hfill \today
\begin{center}
\vskip 0.5cm

{\Large \bf {Characterizing 4-string contact interaction \\
		using machine learning}
}

\vskip 0.5cm

\vskip 1.0cm
{\large {Harold Erbin$^{1,2,3} $ and Atakan Hilmi Fırat$^{1,2}$}}

{\em  \hskip -.1truecm
	$^1$
	Center for Theoretical Physics \\
	Massachusetts Institute of Technology\\
	Cambridge MA 02139, USA
	\\
	\medskip
	$^2$
	NSF AI Institute for Artificial Intelligence and Fundamental Interactions
	\\
	\medskip
	$^3$
	Université Paris Saclay, CEA, LIST\\
	Gif-sur-Yvette, F-91191, France
	\\
	\medskip
	\tt \href{mailto:erbin@mit.edu}{erbin@mit.edu}, \href{mailto:firat@mit.edu}{firat@mit.edu} \vskip 5pt }

\vskip 3.0cm
{\bf Abstract}

\end{center}
\vskip 0.5cm
\noindent
\begin{narrower}
\baselineskip15pt
The geometry of 4-string contact interaction of closed string field theory is characterized using machine learning. We obtain Strebel quadratic differentials on 4-punctured spheres as a neural network by performing unsupervised learning with a custom-built loss function. This allows us to solve for local coordinates and compute their associated mapping radii numerically. We also train a neural network distinguishing vertex from Feynman region. As a check, 4-tachyon contact term in the tachyon potential is computed and a good agreement with the results in the literature is observed. We argue that our algorithm is manifestly independent of number of punctures and scaling it to characterize the geometry of $n$-string contact interaction is feasible.
\end{narrower}
\end{titlepage}

\tableofcontents

\section{Introduction}

Despite possibly being one of the most intricate quantum field theory, our practical knowledge of closed string field theory (CSFT) and its possible solutions, yet alone our understanding of its quantum effects, is still limited after almost three decades from its initial formulation (for reviews see~\cite{Zwiebach:1992ie,deLacroix:2017lif, Erler:2019loq, Erbin:2021smf}). Although there were some attempts towards understanding the classical tachyon potential in the case of bosonic CSFT in the past~\cite{Belopolsky:1994sk, Belopolsky:1994bj, Okawa:2004rh, Moeller:2004yy, Yang:2005iua, Yang:2005ep, Yang:2005rw, Yang:2005rx, Moeller:2006cw, Moeller:2006cv, Moeller:2007mu, Moeller:2008tb, Schlechter:2019hrb}, the closed string tachyon vacuum (or lack thereof) hasn't revealed itself entirely yet. This is primarily due to CSFT being a non-polynomial theory. For the lack of better formulation and the absence of analytical techniques that can overcome these difficulties,~\footnote{Although see the recent developments in hyperbolic string field theory~\cite{Moosavian:2017qsp, Moosavian:2017sev, Costello:2019fuh, Cho:2019anu, Firat:2021ukc, Wang:2021aog, Ishibashi:2022qcz}.} the most straightforward way to progress appears to truncate the theory to some order/level and perform numerical computations. It is possible such approach will yield results sufficiently close to the results of the full theory, just like it does in open string field theory~\cite{Sen:1998sm, Sen:1999xm, Sen:1999nx, Moeller:2000xv, Berkovits:2000hf, Moeller:2000jy, Sen:2000hx, Rastelli:2001jb, Taylor:2002fy, Gaiotto:2002wy}.

In the past, Nicolas Moeller has taken this approach to CSFT first by truncating classical bosonic CSFT (in the minimal-area parametrization) up to quartic order~\cite{Moeller:2004yy}, then up to quintic order~\cite{Moeller:2006cw, Moeller:2006cv, Moeller:2007mu}, and he numerically calculated the tachyon potential and its minimum by level truncation. Even though there has been some progress this way, the fate of closed string tachyon condensation in the bosonic CSFT remains unclear; indicating including terms from higher orders/levels (and understanding their interplay) may be necessary to produce precise results.

In order to truncate the classical bosonic CSFT up to sextic and higher orders, it is necessary to solve the geometry of string contact interactions on six-and higher-punctured spheres. This involves finding Strebel quadratic differentials,~\footnote{Quadratic differentials are reviewed in section~\ref{sec:Rev}. For comprehensive mathematical account see~\cite{strebel1984quadratic}.} obtaining their associated local coordinates to calculate off-shell string amplitudes and finally finding the relevant sub-region (so-called \textit{vertex region}) in the moduli space of the punctured spheres where the moduli integration has to be performed.

Performing all of these numerically was feasible for four-and five-punctured spheres using classical numerical methods, such as Newton's method. However, they seem to fall short and become unfeasible when there are six or more punctures. The basic roadblock is that the equations one needs to solve begin depending on the shape of the so-called critical graph of Strebel differential, which is impossible to obtain without knowing the Strebel differential itself. This informs us that the numerical methods have to be modified in a way that the algorithm should produce Strebel differentials without referring their critical graphs a priori.

In this paper we precisely do this by representing Strebel differentials on four-punctured spheres as an \textit{artificial neural network}. We obtain such network by performing \textit{unsupervised learning} using a custom-built loss function which gets minimized when a quadratic differential is Strebel. Such loss function is, by construction, agnostic of the critical graph and this is the reason it overcomes the hurdle mentioned above. Machine learning algorithms, such as the one we use, have been already found its place in string theory, from exploring the landscape of string theory~\cite{Ruehle:2017mzq, Halverson:2019tkf, Larfors:2020ugo, Ashmore:2021qdf} to obtaining information regarding Calabi-Yau manifolds\cite{Douglas:2020hpv, Anderson:2020hux, Erbin:2020srm, Erbin:2020tks, Erbin:2021hmx, Ashmore:2021ohf, Larfors:2021pbb, Larfors:2022nep, Jejjala:2022lxh} and its overall relation to quantum field theory~\cite{Halverson:2020trp, Halverson:2021aot, Erbin:2021kqf}. In particular, we would like to point out that our algorithm has been partially inspired by the methods described in~\cite{Larfors:2021pbb,Larfors:2022nep}. For a recent review on the applications of data science to string theory see~\cite{Ruehle:2020jrk}.

We note that four-punctured sphere is the first case leading to a non-trivial string contact interaction and reemphasize it has been already solved by Moeller in~\cite{Moeller:2004yy}. Here we just use it as a test ground for our ideas. Even so, since we obtain Strebel differentials as a neural network, our approach is an improvement as once the network is properly trained it can be used to find the Strebel differential for any four-punctured sphere practically \textit{immediately}. It is also philosophically different from Moeller's approach as we are solving for the function itself, instead of just finding the solutions at specific moduli. While one can use a (polynomial) fit to approximate the function, neural networks are more flexible and expressive. In particular, they may be used for non-parametric regression~\cite{SchmidtHieber:2020:NonparametricRegressionUsing,Zhang:2022:DeepLearningMeets} and they can extrapolate outside the training region~\cite{Balestriero:2021:LearningHighDimension,Xu:2021:HowNeuralNetworks}. Let us further note that Moeller stressed that he did not succeed in finding a simple fit in the case of 5-punctured spheres~\cite{Moeller:2006cw}.

We make sure every step of our algorithm is manifestly independent of the number of punctures. In subsequent work~\cite{upcoming_work}, we plan to characterize the string contact interactions for higher-punctured spheres, where benefits of constructing the algorithm this fashion would be apparent. Once the Strebel differential is obtained, we can find the local coordinates by expanding it around the punctures. This alone doesn't specify the so-called mapping radii, but one can easily solve it for by numerically evaluating a specific integral~\cite{Belopolsky:1994sk, Moeller:2004yy}. In this step we make an observation turning this calculation independent of the critical graph as well.

On top of the local coordinates, we also need to solve for the region $\V_{0,n} \subset \M_{0,n}$ where the moduli integration has to be performed. Here $ \M_{0,n}$ is the moduli space of $n$-punctured spheres, while $\V_{0,n}$ is so-called \textit{vertex region}, implicitly determined by taking the lengths of \textit{all} non-contractible curves greater than or equal to certain value in the metric associated with Strebel differential~\cite{Zwiebach:1992ie, Zwiebach:1990ni, Zwiebach:1990nh, Saadi:1989tb, Kugo:1989aa}, which we take $2 \pi$ by convention.\footnote{We only focus on surfaces of genus $0$, i.e. punctured spheres, so that this definition for vertex region can be used. In other words we only consider \textit{classical} CSFT.} These lengths can be computed given Strebel differential and once computed, we can generate a data set to train a neural network to distinguish punctured-spheres in $\V_{0,n}$ from those outside. That is, we can train a network for \textit{the indicator function} that outputs $1$ if the surface is part of the vertex region and $0$ otherwise. In this work we obtain the indicator function $\Theta_{0,4}: \M_{0,4} \simeq \mathbb{C} \setminus \{0,1\}  \to \{0,1\} $ defined by,
\begin{align} \label{eq:IndFunc}
	\Theta_{0,4}(\xi, \xi^\ast) =
	\begin{cases}
	1 \quad \text{if} \quad \xi \in \V_{0,4}\\
	0 \quad \text{if} \quad \xi \notin \V_{0,4}
	\end{cases},
\end{align}
as a neural network. Here $\xi$ denotes the moduli. This allows us to replace
\begin{align} \label{eq:IndFuncC}
	\int_{\V_{0,4}} \to \int_{\M_{0,4}} \Theta_{0,4}(\xi, \xi^\ast) \, ,
\end{align}
and it simplifies the moduli integration in practical terms by eliminating the need for describing the region $\V_{0,4}$ explicitly. We argue an analogous construction for the indicator function would work for higher-punctured spheres and it would be particularly superior, especially in the view of Monte-Carlo integration may be required for higher dimensional moduli integrals.

Lastly, we test our algorithm by computing the coefficient of the 4-tachyon contact interaction. In the conventions of~\cite{Moeller:2004yy}, we report this value to be $v_4 = 72.396$. Comparing with values obtained based on various different techniques, we observe that our results are consistent with those in the literature, see table~\ref{tab:v4} for summary. This supports the validity of our algorithm.
\begin{table}[h]
	\centering
	\begin{tabular}{ | c | c |}
		\hline
		Average NN (Trapezoid) & 72.320 $\pm$ 0.146 \\
		\hline
		Best NN (Trapezoid) & 72.396 \\
		\hline
		Best NN (Monte-Carlo) & 72.366 $\pm$ 0.096 \\
		\hline
		Belopolsky (1994)~\cite{Belopolsky:1994bj}  & 72.39 \\
		\hline
		Moeller (2004)~\cite{Moeller:2004yy} & 72.390 $\pm$ 0.003 \\
		\hline
		Yang \& Zwiebach (2005)~\cite{Yang:2005rx} & 72.414\\
		\hline
	\end{tabular}
	\caption{\label{tab:v4}Values for the 4-tachyon contact interaction $v_4$. The first line (``Average NN'') represents the value obtained by 4 networks (of each type) and computing the integral using the trapezoid method.
	The second and third lines (``Best NN'') show the value for the best result using the trapezoid and Monte Carlo methods. We also add the results in the literature obtained by various authors. The computations are explained in section~\ref{sec:ML}.}
\end{table}

The rest of the paper is organized as follows. In section~\ref{sec:Rev}, we review Strebel differentials, local coordinates, mapping radii, and the indicator function. In particular, we analytically solve Strebel differentials on 4-punctured spheres when all punctures are real and introduce a loss function for Strebel differentials which forms the central part of our algorithm. In section~\ref{sec:ML}, we describe the specifics of our neural networks and their training. In particular we show that the trained network has learned the relevant symmetries of the Strebel differential and produced the analytic results correctly. Moreover, we show that our results are consistent with the fits provided by Moeller in~\cite{Moeller:2004yy}. In the last section we conclude our paper and discuss possible future directions, especially we argue that scaling the algorithm to higher-punctured spheres is expected to be feasible. In appendix~\ref{sec:App} and~\ref{sec:AppB}, we provide some details on numerical evaluations in our work.

\section{The geometry of string contact interaction} \label{sec:Rev}

In this section we review Strebel differentials, local coordinates, mapping radius, and the indicator function. For more details, reader can refer to~\cite{strebel1984quadratic, Belopolsky:1994bj, Belopolsky:1994sk, Moeller:2004yy}. The novel features in this sections are complete analytic characterization of Strebel differentials for 4-punctured sphere when all punctures lie on a great circle and introduction of a loss function for Strebel differentials as well as few simplifying observations on the calculation of mapping radii.

\subsection{Strebel quadratic differential}

Imagine a $n$-punctured sphere $\Sigma_{0, n}$, with punctures placed at $P = \{\xi_1, \dots, \xi_n \}$ assuming none of the punctures are at infinity. We always fix the positions of the last three punctures to pre-determined positions by appropriate PSL(2,$\Comp$) transformations. We are interested in quadratic differentials that have a double pole at each punctures with residue equal to $-1$.~\footnote{The residues can be chosen different from each other in principle as long as they are still negative, but this situation is not relevant for the current formulation of CSFT.} In general they can be written as
\begin{align} \label{eq:quaddiff}
	\vf = \f (z) dz^2 = \sum_{i=1}^n
	\left[
	{-1 \over (z-\xi_i)^2} + {c_i \over z-\xi_i}
	\right] dz^2 \, ,
\end{align}
where $c_i \in \Comp$, $i=1, \dots n$ are a priori undetermined variables which we call \textit{accessory parameters}. These are unconstrained and they are only going to be fixed upon demanding~\eqref{eq:quaddiff} to be a special type of quadratic differential, Strebel differential. The double pole structure with residues equal to $-1$ can be argued by demanding the metric associated with quadratic differential $d s = \sqrt{ |\f(z)| } |dz|$ (so-called $\vf-$metric) is of the flat cylinder of circumference $2 \pi$ when sufficiently close to a puncture~\cite{strebel1984quadratic}. The flat cylinders here corresponds to external strings.

Given we have the punctures at $P = \{\xi_1, \dots, \xi_n \}$, the point at infinity $z = \infty$ has to be regular in general for string contact interactions. Inverting the coordinates by $w = 1/z$ it is easy to see that the quadratic differential $\vf$ takes the following form around $z = \infty$ (or $w = 0 $)
\begin{align} \label{eq:quadinfexp}
	\vf
	= {1 \over w^4} \f \left( {1 \over w} \right) dw^2
	= \sum_{i=1}^n \left[
	 {c_i \over w^3} + {-1 + c_i \xi_i \over w^2} + {-2\xi_i + c_i \xi_i^2 \over w } + \mathcal{O}(w^0)
	\right] dw^2 \, .
\end{align}
This leads to following three linear conditions among accessory parameters $c_i$
\begin{align} \label{eq:QuadCond}
	\sum_{i=1}^n c_i = 0 \, ,
	\quad
	\sum_{i=1}^n (-1 + c_i \xi_i) = 0 \, ,
	\quad
	\sum_{i=1}^n (-2\xi_i + c_i \xi_i^2) = 0 \, .
\end{align}
Notice this still leaves us with $n-3$ undetermined accessory parameters $c_i$. Also notice this explains why we haven't included regular terms in~\eqref{eq:quaddiff}: it just leads to more singular terms around $z = \infty$.

Notice there are  no undetermined accessory parameters when $n=3$. If we place the punctures at $P = \{ 0, 1, \infty \}$, we find the quadratic differential~\eqref{eq:quaddiff} to be
\begin{align}
\f(z) =
\left[
{-1 \over z^2} + {-1 \over (z-1)^2}
+ {-1 \over z} + {1 \over z-1}
\right]
= - {z^2 - z  + 1 \over z^2 (z-1)^2 }
= - { ( z- (-1)^{1/3} )  ( z - (-1)^{-1/3} ) \over z^2 (z-1)^2}  \, .
\end{align}
It is well known that such differential leads to Witten's vertex for closed strings~\cite{witten1986non, Sonoda:1989wa}.

\begin{figure}[h]
	\centering
	\includegraphics[height=7cm, width=7cm]{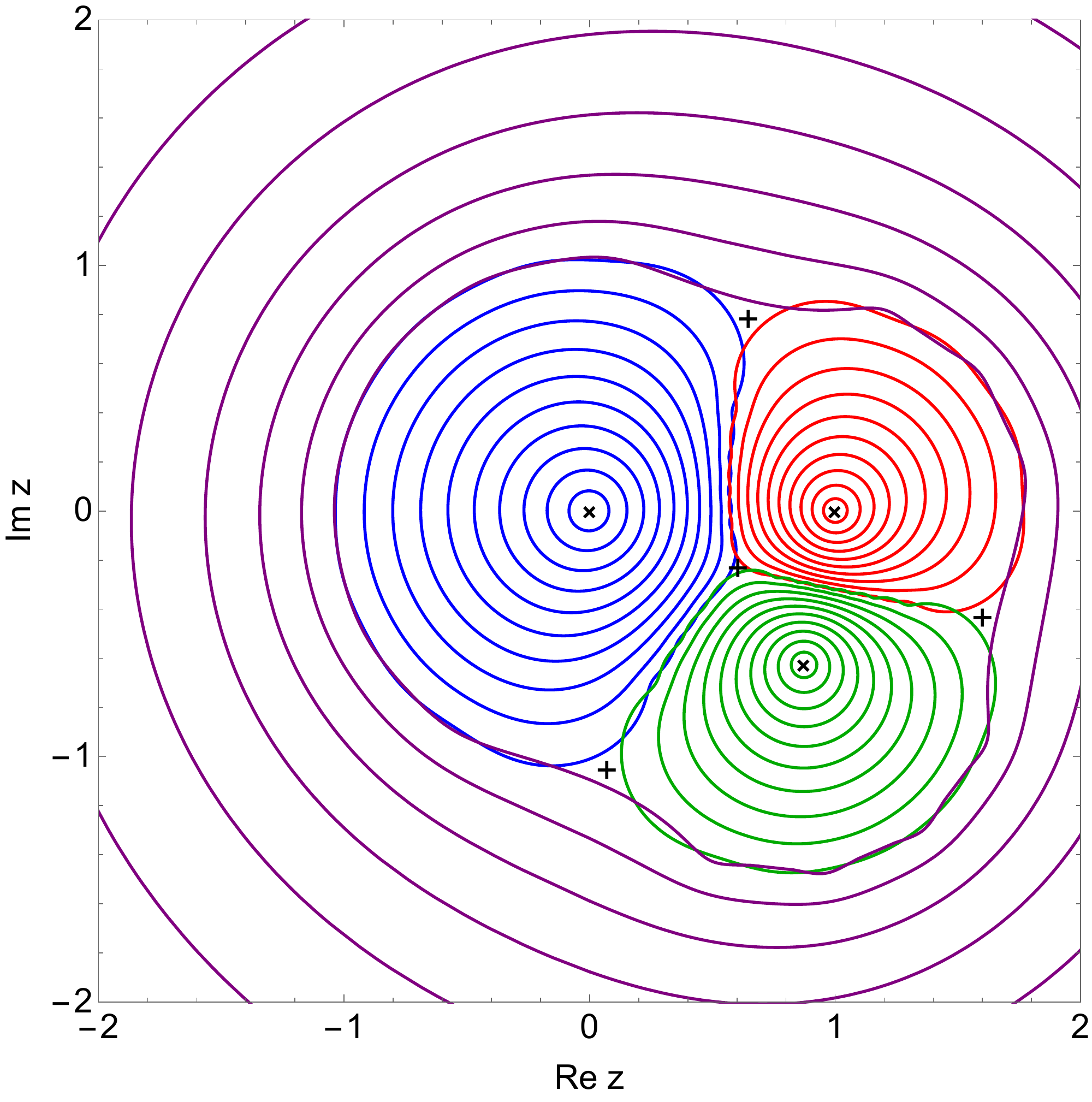}
	\hspace{1cm}
	\includegraphics[height=7cm, width=7cm]{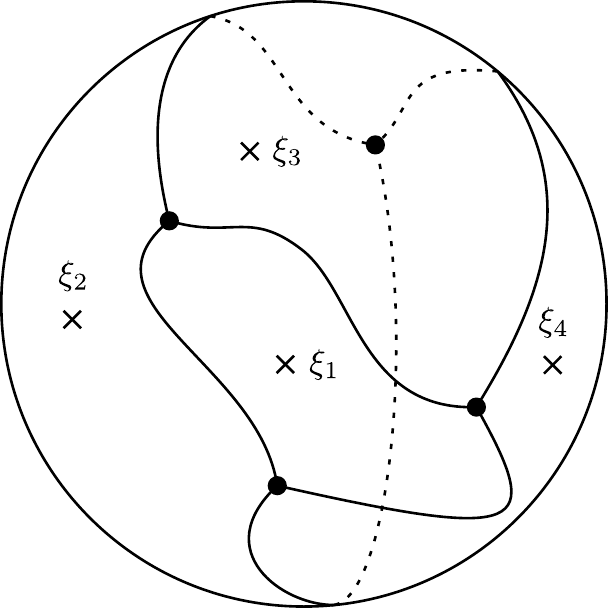}
	\caption{\label{fig:example}The trajectory structure of the Strebel differential when punctures are at~$P = \{0, 1, 0.8734-0.6242i, \infty\}$ (left). We marked the positions of punctures and zeros by crosses and plusses respectively. The inaccuracies around the zeros are due to evaluating the trajectories as an expansion after~\eqref{eq:LocCord}. The critical graph is a tetrahedron whose sketch on $\mathbb{CP}^1$ is shown on the right.}
\end{figure}

Before we introduce Strebel differentials, let us introduce some nomenclature. Define \textit{horizontal trajectory} as a path such that $\vf > 0 $ along it. A \textit{critical trajectory} is a horizontal trajectory that begins and ends on a zero of $\vf$. It is easy to argue $n+2$ critical trajectory would emanate from $n$-th order zero and the orders of zeros would add up to $2n-4$~\cite{strebel1984quadratic}. The union of critical trajectories of $\vf$, together with their endpoints, is called \textit{critical graph}. \textit{Strebel differential} is then defined as a quadratic differential with double poles of residue $-1$ whose critical graph forms a non-empty measure zero set and is connected. Horizontal trajectories of such a differential foliate the entire surface~\cite{strebel1984quadratic}. An example of the trajectory structure of a Strebel differential is shown in figure~\ref{fig:example}.

The Strebel differential exists and is unique for every punctured sphere (see Theorem 23.5 in~\cite{strebel1984quadratic}). From this, and the fact that double poles with negative residues give closed horizontal trajectories sufficiently close to the punctures, there exists a set of accessory parameters, unique up to relations given in~\eqref{eq:QuadCond}, such that the quadratic differential~\eqref{eq:quaddiff} is Strebel given punctures at $P$. Our goal is to find such accessory parameters $c_i$ as a function of the position of punctures.

We remark that $\vf-$metric associated with Strebel differential $\vf$ is the metric of minimal-area and it looks like flat cylinders grafted to each other dictated by the critical graph of $\vf$~\cite{Zwiebach:1990ni, Zwiebach:1992ie, strebel1984quadratic, Zwiebach:1990nh}. However, CSFT in minimal-area parametrization actually demands solving for minimal-area metrics for which the lengths of \textit{all} non-contractible curves are greater than or equal to $2 \pi$ for consistency. While the latter condition certainly fails for Strebel differentials when surfaces are sufficiently close to degeneration, for surfaces that are part of the classical elementary interaction (that is, those in the vertex region $\mathcal{V}_{0, n}$) this condition is satisfied by definition. So after finding Strebel differential everywhere on the moduli space $\mathcal{M}_{0, n}$ (\textit{or} in any region containing  $\mathcal{V}_{0, n}$), it is possible to map out the vertex region $\mathcal{V}_{0,n}$ by checking the lengths of non-contractible curves. Furthermore, once Strebel differential is known it is possible to find the local coordinates characterizing the geometry of $n$-string contact interaction. We explain these in subsequent subsections.

We finally note that for the quadratic differentials on the surfaces outside the vertex region, that is, those in the so-called \textit{Feynman region} $\mathcal{F}_{0, n} = \mathcal{M}_{0, n} \setminus \mathcal{V}_{0, n}$, Strebel differential is not the right type of quadratic differential from the perspective of CSFT: we have to use a quadratic differential of the form~\eqref{eq:quaddiff} whose associated metric is of minimal-area under the condition that the lengths of all of its non-contractible curves is greater than or equal to $2 \pi$.~\footnote{Even though Strebel differentials in Feynman region do not seem to be relevant for CSFT, we would like to point out that it has found applications in worldsheet approaches to AdS/CFT correspondence, see~\cite{Gopakumar:2005fx, Gaberdiel:2020ycd, Bhat:2021dez, Knighton:2022ipy}.} We call such differentials \textit{Zwiebach differentials} (see Theorem 3.2 in~\cite{Zwiebach:1990nh}). In the vertex region $\mathcal{V}_{0,n}$ the definition of Zwiebach differentials coincides with the definition of Strebel differential, but in the Feynman region $\mathcal{F}_{0, n}$ they differ: the critical graph of Zwiebach differential becomes disconnected~\cite{Zwiebach:1990nh}. Geometrically this corresponds having internal cylinders corresponding to string propagators. In fact, Zwiebach differentials are examples of more general type of differentials called \textit{Jenkins-Strebel (JS) differential} for which the critical graph forms a non-empty measure zero set but not necessarily connected. Zwiebach differentials can be shown to exist and be unique (see Theorem 5.1 in~\cite{Zwiebach:1990nh}). We emphasize that the accessory parameters for Strebel and Zwiebach differentials are distinct functions of the moduli in the Feynman region $\mathcal{F}_{0, n}$. In this study, we only consider Strebel differentials.~\footnote{Although see section~\ref{sec:Con} for the discussion on how ideas here can be extended to obtain Zwiebach differential as a neural network.}

\subsection{Complex length and loss function}

It is hard to work with the definition of Strebel differential given in previous subsection. Here we provide an equivalent characterization more amenable to analytical and numerical investigations. Begin with defining \textit{the complex length} between zeros $z_i, z_j$ of $\vf$ as
\begin{align} \label{eq:CompLen0}
	\ell (z_i, z_j) \equiv \int_{z_i}^{z_j} \sqrt{\phi(z)} dz \, .
\end{align}
The path of integration here is chosen so that it avoids any branch cuts. Since the branch structure of $\sqrt{\phi(z)} $ is hard to keep track numerically, we are going to replace the square root with the \textit{continuous square root} $ \sqrt[\pm]{} $, just like in~\cite{Moeller:2004yy}. That is, we are going to define the domain of square root in the double cover of complex plane sans the origin, where it is holomorphic. This would make the overall sign of the continuous square root, hence complex length, ambiguous. But, as we shall see, the problems of this sort would be of technical natural and they can be easily overcome. More details on numerical evaluation of continuous square root is given in appendix~\ref{sec:App}.

With such replacement the complex length is now taken to be
\begin{align} \label{eq:CompLen}
	\ell (z_i, z_j) \equiv \int_{z_i}^{z_j} \sqrt[\pm]{\phi(z)} dz \, .
\end{align}
Assuming the integrand doesn't vanish on the path of integration and the path is non self-intersecting, the integrand is holomorphic around some neighborhood of the path and the integral is equal to~\eqref{eq:CompLen0} up to overall sign (with appropriate choice of branch cut for $\sqrt{}$). Hence we can deform the path of integration for~\eqref{eq:CompLen} freely without changing the value of the integral as long as the endpoints are fixed, the path doesn't cross any punctures and/or intersects itself.
Therefore, for convenience, we evaluate the integral in~\eqref{eq:CompLen} always on the straight line
\begin{align} \label{eq:IntPath}
z(t) = {z_j + z_i \over 2} + {z_j - z_i \over 2 } \, t \quad
\text{for} \quad
t \in \left[ -1, 1 \right] \, .
\end{align}
The details on numerical evaluations the complex length can be found in appendix~\ref{sec:App}.

But notice, regardless of where the path lies relative to the punctures, the (absolute value of) imaginary part of the complex length would be the same. To see this, just note that deforming the path of integration over the puncture would pick up the residue of $\sqrt[\pm]{\phi(z)}$ and this is always purely imaginary by having a residue equal to $-1$ for the  double poles~\eqref{eq:quaddiff}. This makes the shift real and equal to $2 \pi$, leading to no change for the imaginary part of the complex length.

This reasoning implies the complex length is real for Strebel differentials, as we can deform the path of integration~\eqref{eq:IntPath} to a critical trajectory between $z_i, z_j$, which has $\vf \geq 0$. This makes the integrand equal to the line element $ds = \sqrt{|\phi(z)|} |dz|$ up to sign and manifestly real. We remark that the (absolute value of) complex length may not always give the geodesic distance between zeros $z_i, z_j$ in the $\vf-$metric for Strebel differentials due to the sign ambiguity and the placement of the punctures relative to the path of integration.

Above we essentially provide a necessary condition for a quadratic differential~\eqref{eq:quaddiff} to be Strebel: if $\vf$ is Strebel then the complex length is real between all zeros of $\vf$. In fact the other direction is true as well. That is, if $z_i, z_j$ are zeros of a quadratic differential of the form~\eqref{eq:quaddiff}, $\phi(z_i) = \phi(z_j) = 0$, then we have
\begin{align} \label{eq:AltCond}
	\vf \; \text{is Strebel} \iff  \forall \; z_i, z_j, \;  \mathrm{Im} ( \ell (z_i, z_j) )= 0 \, .
\end{align}
In order to argue for the sufficient condition, we just have to show the integrand of the complex length, $\sqrt[\pm]{\phi(z)} dz$, is real throughout some path among all zeros and then its square, $\phi(z) dz^2$ would define a Strebel differential as it is going to be a single-valued quadratic differential of the form~\eqref{eq:quaddiff} and its critical graph would be measure zero and connected. This is easy to accomplish, as we can deform the path between each zeros to the path that would make the integrand real by beginning from one zero and moving in the direction that sets the imaginary part to zero. Since $\mathrm{Im}  ( \ell (z_i, z_j) )= 0 $ for all $i,j$, we would be guaranteed to hit another zero after this procedure and this makes $\phi(z) dz^2$ a Strebel differential.

The condition~\eqref{eq:AltCond} gives an alternative formulation for Strebel differentials. In fact this is the condition solved by Moeller using Newton's method~\cite{Moeller:2004yy,Moeller:2006cw, Moeller:2007mu}. Observe that the existence and uniqueness of Strebel differentials translates to the existence and uniqueness of the solution of the equations in the right-hand side of~\eqref{eq:AltCond} in terms of accessory parameters up to relations in~\eqref{eq:QuadCond}. Note that there are ${2n-4 \choose 2 }$ distinct equations in the right-hand side of~\eqref{eq:AltCond}. However, it is actually sufficient to demand vanishing of $\dim( \mathcal{M}_{0,n} ) = 2n-6$ imaginary parts of complex lengths by dimensional counting. This shows the set of equations in the right-hand side of~\eqref{eq:AltCond} is in fact over-determined.

Now, define the following function of quadratic differentials of the form~\eqref{eq:quaddiff} motivated by the conditions in the right-hand side of~\eqref{eq:AltCond}
\begin{align} \label{eq:CostFunc}
\mathcal{L}_{0,n} (\vf) \equiv  {2n-4 \choose 2}^{-1} \sum_{i< j}  ( \mathrm{Im} \; \ell (z_i, z_j) )^2 \, .
\end{align}
Here $i,j = 1, \dots, 2n-4$ runs over the zeros of $\vf$ (accounting degeneracy) and the overall factor is for normalization. We call this function \textit{loss function} for reasons that is going to be apparent in section~\ref{sec:ML}. Observe this function can be \textit{unambiguously} evaluated using the integral~\eqref{eq:CompLen} and taking the path of integration to be the straight line~\eqref{eq:IntPath}: as we square each imaginary part and their sign ambiguity becomes irrelevant. By construction we have $\mathcal{L}_n \geq 0$.

The loss function~\eqref{eq:CostFunc} has a \textit{unique} global minimum as a function of accessory parameters $c_i$ (up to relations~\eqref{eq:QuadCond}) at fixed positions of punctures $\xi_i$ by the existence and uniqueness of Strebel differentials and its value is equal to zero. So, it is in principle possible to obtain Strebel differentials by minimizing the loss function~\eqref{eq:CostFunc} in the space of accessory parameters given the positions of punctures. This optimization problem is perfectly suited to machine learning and it is how we are going to approach finding Strebel differentials in section~\ref{sec:ML}. In particular, the advantage of this approach is clear from the fact that the loss function constructed out of~\eqref{eq:quaddiff} is totally agnostic of the shape of critical graphs, which made the previous approaches to solving Strebel differential slightly convoluted as we mentioned.

It is an interesting question whether the loss function~\eqref{eq:CostFunc} has another extremum. Our experimental investigation in the case of 4-punctured sphere informs us that even if there is, it hasn't made an appearance in our algorithm. So we \textit{assume} there is no other extremum of the loss function~\eqref{eq:CostFunc} for all intents and purposes. It may be interesting to rigorously establish this is the case.

\subsection{Strebel differential on 4-punctured sphere}

Since we are going to test our algorithm for 4-string contact interaction, let us focus on Strebel differentials on 4-punctured spheres more. Begin with placing punctures at $P = \{0,1, \xi, \infty \}$ by performing $PSL(2,\mathbb{C})$ transformation. We see $\xi$ here is the moduli. Since there is single accessory parameter after solving the modified version of conditions in~\eqref{eq:quadinfexp} when one of the punctures is at $z=\infty$, it can be shown that the quadratic differential~\eqref{eq:QuadCond} can be put into the following form:
\begin{align} \label{eq:quad4}
	\f(z)
	&= {-z^4 + a z^3 + (2 \xi - (1+ \xi) a ) z^2 + a \xi z -\xi^2 \over z^2 (z-1)^2 (z- \xi)^2 }
	&=-  {  (z-z_1) (z-z_2) (z-z_3) (z-z_4) \over z^2 (z-1)^2 (z- \xi)^2} \, ,
\end{align}
where $a = a (\xi, \xi^\ast)$ is the single accessory parameter and $z_i$, $i=1, 2,3,4$ are the zeros of the quadratic differential.~\footnote{In this work we used analytical expressions for the zeros of quadratic differentials. Obviously, this wouldn't work in higher-punctured sphere, but we think this is just a technical issue which we aim to confront in our future work~\cite{upcoming_work}.} As we have emphasized earlier, finding Strebel differential is equivalent to finding the function $a = a (\xi, \xi^\ast)$.

There are certain symmetries the accessory parameter $a = a (\xi, \xi^\ast)$ enjoys. These are:
\begin{itemize}
	\item \underline{The involution symmetry.} The complex conjugate of Strebel differential on the surface $\S_{0,4}$ would be the Strebel differential for the conjugated Riemann surface $\S_{0,4}^\ast$. That is, if the accessory parameter corresponding to the moduli $\xi$ is $a$, then the accessory parameter corresponding to the moduli $\xi^\ast$ is $a^\ast$. In particular this shows for $\xi \in \Real$, we have $a \in \Real$. Clearly, similar symmetry holds for $n$-punctured spheres.

	\item \underline{$PSL(2, \Comp)$ symmetries permuting $\{0,1, \infty \}$.} There are 6 such transformations but they are generated by the following two transformations
	\begin{align} \label{eq:ConfAcc0}
	z \to z' = 1-z \quad
	\text{and} \quad
	z \to z' = {1 \over z} \, .
	\end{align}
	and the position of the puncture at $z=\xi$ as well as the quadratic differential changes accordingly. These transformations shouldn't change a quadratic differential being Strebel, so it can be shown performing~\eqref{eq:ConfAcc0} induces following transformations for the moduli and the accessory parameter of a Strebel differential
	\begin{align} \label{eq:ConfAcc1}
	\xi \to 1 - \xi \implies a \to - a + 4
	\quad \text{and} \quad
	\xi \to {1 \over \xi} \implies a \to { a \over \xi} \, .
	\end{align}
	These symmetries can be generalized to higher-punctured spheres in an obvious fashion.

\end{itemize}

We can use these symmetries to solve for the accessory parameter for certain values of $\xi$. For example, for $\xi = 1/2 + (\sqrt{3}/2) i = e^{i \pi/3}$ we have $1/\xi = 1-\xi$, which shows
\begin{align} \label{eq:exacta}
-a + 4 = {a \over \xi}
\implies
a = {4 \xi \over 1 + \xi} = 2 + {2 i \over \sqrt{3}} \, .
\end{align}
In fact the symmetries fix the critical graph to be a regular tetrahedron whose sides have lengths equal to $2 \pi/3$ in this case~\cite{Belopolsky:1994bj}. We can also find the Strebel differential when $\xi = 1/2$. The moduli satisfies $\xi = 1- \xi$ so we have $a = -a + 4 $, which shows $a=2$. This subsequently shows $a=4, 0$ for $\xi = 2, -1$ respectively. Furthermore, it is actually possible to find the Strebel differentials for the moduli between $0 < \xi <1$. Our claim is $a = 4 \xi$ for $0 < \xi <1$.

In order to argue for this, recall that the critical graph of a Strebel differential on 4-punctured sphere has to be topologically planar tetrahedron on $z-$plane in general~\cite{Saadi:1989tb, Belopolsky:1994bj}--one example has been already shown in figure~\ref{fig:example}. Now recall that the critical graph in the complex conjugated surface has to be the mirror image of the original graph around the real axis by the involution symmetry. Combining these two facts and taking $\xi \in \mathbb{R}$, we see the mirror image of the critical graph has to be itself. However for planar tetrahedral graphs this is impossible: the graph should degenerate. That means at least two of the zeros coincide and the other two zeros coincide as well by the involution symmetry. Pair of double zeros zero emanates $4$ critical trajectories each now.

In the language of quadratic differentials that means we now have a pair of double zeros that are complex conjugates of each other. That is, Strebel differential has to take the form
\begin{align}
	\vf = - {  (z-z_1)^2  (z- \bar z_1)^2 \over z^2 (z-1)^2 (z- \xi)^2} dz^2 \, ,
\end{align}
when $\xi \in \mathbb{R}$. Comparing with the form in~\eqref{eq:quad4}, solving for $a$, and using the fact that $ a = 2$ when $\xi = 1/2$ it can be shown that $a=4\xi$ for $ 0 < \xi < 1$ after some algebra. Using the symmetries in~\eqref{eq:ConfAcc1} we can further show $a =4$ for $\xi > 1$ and $a=0$ for $\xi <0$.

The situation when punctures collide is more subtle. Demanding continuity of $a$ suggests we should have $a=0,4$ for $\xi = 0,1$. Indeed, if this is the case, we see the quadratic differentials become
\begin{align}
	\vf = -{dz^2 \over (z-1)^2}
	\quad \text{and} \quad
	\vf = - {dz^2 \over z^2} \, .
\end{align}
for $\xi = 0, 1$ respectively. But notice these are the quadratic differentials describing an infinite flat cylinders whose punctures are at $z=1,\infty$ and $z=0, \infty$ respectively. This is exactly what it should be expected from the degeneration of Strebel differentials when we take $\xi \to 0,1$: the residue condition forces having a single cylinder. The situation at $\xi = \infty$ is similar to $\xi =0$, only difference being that we have to perform this calculation after inversion $z \to 1/z$. In particular, it can be shown that the limit doesn't depend on which value of $a$ we use (0 or 4). Summarizing, we find the following expression for the accessory parameter $a$ when the moduli is real:
\begin{align} \label{eq:PlanarAcc}
	a(\xi = \xi^\ast) =
	\begin{cases}
	0 & \xi\leq 0 \\
	4 \xi & 0\leq \xi \leq 1 \\
	4 & 1 \leq \xi
	\end{cases} \, .
\end{align}
The plot of this function is shown in figure~\ref{fig:a}.

With the accessory parameter is available for Strebel differentials when $\xi \in \mathbb{R}$, we can find the lengths of the sides of the critical graph as a function of the moduli.  In order to do that recall the length of the geodesics homotopic to a puncture is always equal to $2 \pi$. Then, since the critical graph is degenerated in the way described above, we only need to calculate the length of a single side, $\ell$, and the lengths of the other sides would be either equal to this or $2 \pi - \ell$. We can find $\ell$ by carefully evaluating~\eqref{eq:CompLen} using~\eqref{eq:quad4} with the accessory parameter $a$ given in~\eqref{eq:PlanarAcc} for $\xi \in \mathbb{R}$. The result is
\begin{align}
\ell \, (\xi = \xi^\ast) =
\begin{cases}
2 \pi - 4 \arctan (\sqrt{-\xi}) & \xi\leq 0 \\
4 \arctan \left( \sqrt{-1 + {1 \over \xi}} \right)  & 0\leq \xi \leq 1 \\
4 \arctan (\sqrt{\xi -1}) & 1 \leq \xi
\end{cases}
\quad \text{or} \quad
\xi =
\begin{cases}
- \cot^2 \left( {\ell \over 4} \right)  & \xi\leq 0 \\
\cos^2 \left( {\ell \over 4} \right)   & 0\leq \xi \leq 1 \\
\sec^2 \left( {\ell \over 4} \right)  & 1 \leq \xi
\end{cases}
\, .
\end{align}
It is clear that the limits make sense from this expression. For example for $\xi = 1/2$, we have $\ell = \pi$ and this can be alternatively argued by the symmetry of this case. In fact, notice $2 \ell$ and $4 \pi - 2 \ell$ are the lengths of the non-contractible geodesics non-homotopic to punctures and they are
equal to $2 \pi$ only when  $\xi = -1, 1/2, 2$ and less than $2 \pi$ otherwise. In other words, these are the only real moduli that are in the vertex region $\V_{0,4}$. This result is consistent with~\cite{Saadi:1989tb}.

Before closing off this subsection, we note that our argument for degenerate Strebel differentials on four-punctured spheres may not generalize to \textit{every} type of degeneration of higher-punctured spheres. This is mostly due to the symmetry of the critical graph when all (or some) moduli taken to be real may not be as restrictive as it does for the four-punctured spheres. Still, it may be possible to find exact solutions for specific type of degenerations. Since we don't need them immediately, we plan to investigate the degeneration behavior in more detail in our upcoming work~\cite{upcoming_work}.

\subsection{Local coordinates, mapping radii, and vertex region}

Calculating off-shell string amplitudes on any Riemann surface requires a choice of local coordinates up to an overall phase around the punctures~\cite{Zwiebach:1992ie}. Our case of interest, the local coordinates for the $n$-string contact interactions, can be obtained using Strebel differential on $n$-punctured spheres as they are described through how $n$ flat semi-infinite cylinders are grafted at the critical graph of Strebel differential~\cite{Belopolsky:1994sk,Moeller:2004yy, Zwiebach:1990ni}. Following the conventions of~\cite{Moeller:2004yy}, this means one can find $n$ analytic maps $h_i$ of the form $(i = 1, 2, \cdots, n)$
\begin{align} \label{eq:LocCord}
	z = h_i ( w_i ) &= \xi_i + \rho_i w_i + \sum_{\alpha = 2}^{\infty} d_{i, \alpha-1} ( \rho_i w_i)^{\alpha} \\
	&= \xi_i + (\rho_i w_i ) + d_{i,1} (\rho_i w_i )^2 + d_{i,2} (\rho_i w_i )^3 + \cdots,
	\quad \text{with} \quad
	\rho_i, \; d_{i, \alpha} \in \mathbb{C}, \; \alpha = 1, 2, \cdots \, , \nonumber
\end{align}
from the punctured disks $D_i = \{ w_i \in \mathbb{C} \; | \; 0< |w_i| < 1 \}$ to $n$-punctured sphere for which the Strebel differential takes the form it takes for the flat cylinders in $w_i$ coordinates
\begin{align} \label{eq:CycQD}
	\vf = - {d w_i^2 \over w_i^2} \, ,
\end{align}
and the unit circles $|w_i| = 1$ are mapped to its critical graph. Here $w_i$ would be the local coordinates for the string contact interactions (sometimes called \textit{natural coordinates}) for which vertex operators are inserted. It can be shown that such coordinates always exist~\cite{strebel1984quadratic}.

Notice how we have organized the expansion~\eqref{eq:LocCord}. This was because of convenience: as the overall phase of the local coordinates is irrelevant for CSFT we chose $\rho_i \in \mathbb{R}$ without loss of generality and we defined the rest of the coefficients accordingly. Here $\rho_i = |d h_i(w_i) / d w_i|_{w_i = 0}$ is called \textit{mapping radius} associated with the puncture $\xi_i$.

Our primary goal is to find the maps~\eqref{eq:LocCord}, i.e. to find $d_{i,\alpha}$ and $\rho_i$. Except for the mapping radius, the coefficients in the expansion~\eqref{eq:LocCord} can be found by expanding the Strebel differential around the puncture $z = \xi_i$
\begin{align}
	\vf = \left[
	-{1 \over (z-\xi_i)^2 } + {b_{i,-1} \over z-\xi_i } + b_{i,0} + b_{i,1} (z-\xi_i) + \cdots
	\right] dz^2 \, ,
\end{align}
and setting it equal to~\eqref{eq:CycQD}, along with using the expansion for $z = h_i ( w_i )$ in~\eqref{eq:LocCord}. Comparing term by term in $w$, we can solve $d$ coefficients in terms of $b$ coefficients. First few terms are
\begin{subequations}
\begin{align}
	d_{i,1} &= {1 \over 2} b_{i,-1} \, , \\
	d_{i,2} &= {1 \over 16} (7 b_{i,-1}^2 + 4 b_{i,0}) \, , \\
	d_{i,3} &= {1 \over 48} (23 b_{i,-1}^3 + 28 b_{i,-1} b_{i,0} + 8 b_{i,1}) \, ,
\end{align}
\end{subequations}
Note that the $b$ coefficients, therefore $d$ coefficients, are determined by the accessory parameters $c_i$, so knowing the latter is sufficient to construct the maps $z = h_i ( w_i )$ up to mapping radius. For example, we see $b_{i, -1} = c_i$, hence $d_{i,1} = c_i/2$.

Finding mapping radii associated with the punctures takes more effort. To that end, begin by writing the local coordinate around $z= \xi_i$ by equating~\eqref{eq:quaddiff} and~\eqref{eq:CycQD} as~\cite{Belopolsky:1994sk}
\begin{align} \label{eq:InvLocCo}
	w_i (z)  = \exp \left(
	-i \int^{z}_{z_c} \sqrt[\pm]{\phi(z')} dz'
	\right) \, ,
\end{align}
where $z_c$ is some point on the critical trajectory \textit{surrounding} the puncture $z = \xi_i$ and the path of integration here is the straight line from the puncture to $z= z_c$. The lower bound of the integral makes sure $|w_i| = 1$ when $z$ lies on the critical trajectory surrounding the puncture, as the integral just evaluates to real number in this case by repeating the arguments made below~\eqref{eq:IntPath}. We implicitly adjust the choice of sign of the exponent to guarantee $|w_i| \leq 1$.

\begin{figure}[h]
	\centering
	\includegraphics[width=3cm]{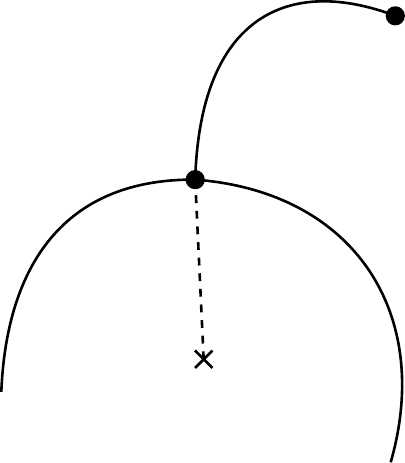}
	\hspace{2cm}
	\includegraphics[width=3cm]{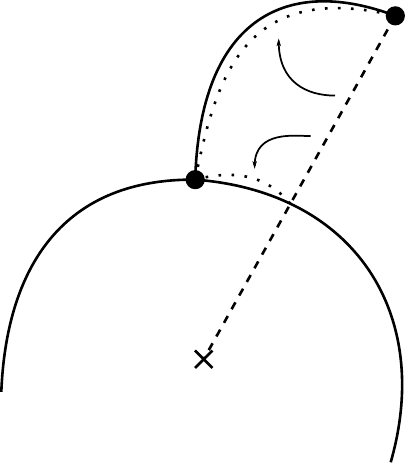}
	\caption{\label{fig:deformation}Part of the path of integration in~\eqref{eq:InvLocCo} can be deformed to the dotted path. The integration over the dotted path produces a real number, resulting in an irrelevant phase for the local coordinates~\eqref{eq:InvLocCo}. In extension, this part also doesn't contribute to the mapping radii~\eqref{eq:map_rad} below.}
\end{figure}

Here we make a small observation that  has apparently gone unnoticed in the literature. We demanded $z_c$ to lie on the critical trajectory \textit{surrounding} the puncture $z = \xi_i$ above, but actually this can be relaxed and one can choose $z_c$ to be lying \textit{anywhere} on the critical graph. To argue for this, let $z_c$ to be any point on the critical graph and notice the integral in the exponent of~\eqref{eq:InvLocCo} can be deformed as shown in figure~\ref{fig:deformation}. But then the contribution to the integral from this ``outside'' part becomes real following a similar argument made below~\eqref{eq:IntPath} and this just results in an irrelevant phase for the local coordinate.  In particular, notice that we can take $z_c$ to be \textit{any} zero of the quadratic differential. With this choice, the dependence of the local coordinates~\eqref{eq:InvLocCo} (and in extension, the mapping radii~\eqref{eq:map_rad}) to the shape of the critical graph drops out.

Using~\eqref{eq:InvLocCo} we can obtain an integral expression for the mapping radii. It is~\cite{Belopolsky:1994sk}
\begin{align} \label{eq:map_rad}
\log \rho_i = \lim_{\epsilon \to 0} \left( \mathrm{Im} \int_{\xi_i + \epsilon}^{z_c} \sqrt{\phi(z')} dz' + \log | \epsilon | \right) \, .
\end{align}
Note that such limit exists with our choice of sign in the exponent of~\eqref{eq:InvLocCo} and assuming $\xi+ \epsilon$ lies on the straight path from the puncture at $z=\xi$ to $z=z_c$. Details on numerical evaluation of such an integral are relegated to appendix~\ref{sec:App}. In passing we note that it is possible to obtain local coordinates for 4-punctured spheres analytically when $\xi \in \mathbb{R}$. However, as we have stated earlier, these surfaces are not relevant from the perspective of CSFT, so we opt out reporting them here.

Calculating off-shell string amplitudes on Riemann surfaces not only requires a choice of local coordinates around the punctures, but also suitable choice of vertex region in the associated moduli space~\cite{Zwiebach:1992ie}. Instead of trying to describe the vertex region explicitly we can consider its associated indicator function. Such function is already defined in the introduction in~\eqref{eq:IndFunc} for the case of 4-punctured spheres. Here we give a general definition for the case of $n$-punctured sphere
\begin{align} \label{eq:IndFunc1}
\Theta_{0,n}(\xi, \xi^\ast) =
\begin{cases}
1 \quad \text{if} \quad \xi \in \V_{0,n}\\
0 \quad \text{if} \quad \xi \notin \V_{0,n}
\end{cases} \, .
\end{align}
The criteria for $\xi \in \V_{0,n}$ is having \textit{all} non-contractible curves in the $\vf$-metric to be greater than or equal to $2 \pi$~\cite{Zwiebach:1992ie, Zwiebach:1990ni, Zwiebach:1990nh, Saadi:1989tb}. For the case of $\vf$-metric this means that it is sufficient to check the lengths of the critical trajectories separating 2 or more punctures from the rest, i.e. geodesics that are not homotopic to a puncture.

These lengths can be computed by  finding the geodesic lengths between each zero of Strebel differential then combining them up suitably. Since the critical graph of a Strebel differential is an undirected graph and we can assign geodesic length to an edge of such graph, it is useful to arrange the associated data into (weighted) adjacency matrix $M$ as follows:
\vspace{0.1cm}
\begin{align}
M_{ij} =
\begin{cases}
\ell_{ij} & \text{when there is an edge of length $\ell_{ij}$ begins at $i$th zero and ends at $j$th zero} \\
0  & \text{otherwise}
\end{cases} .
\end{align}
Here $M$ is a symmetric $(2n-4) \times (2n-4)$ matrix as there are $2n-4$ zeros of $\vf$ (including degeneracy). Elements in the diagonal are zero (hence $M$ is traceless) and each row and column contains at most 3 non-zero elements as zeros of $\vf$ emit 3 critical trajectories for a generic moduli. Notice there are certain nonzero co-dimension loci in the moduli space where some zeros of $\vf$ may coincide. In this case we would set the elements of $M$ corresponding to their connections to zero.

Once such matrix is constructed, it is a simple matter to extract the length of all non-contractible curves: this is what we have done in the case of $n = 4$. But notice for the purposes of the indicator function~\eqref{eq:IndFunc1} we just need to check the length of the shortest non-contractible geodesic and this can be found with relative ease given $M$, such as using Dijkstra’s shortest path algorithm~\cite{dijkstra2022note}.

So the only really contention here is to find the lengths associated with each edge. This can be done by solving the critical trajectories and calculating their lengths. Recall a critical trajectory is a horizontal trajectory begins and ends on a zero of $\vf$. So given a zero, it is possible to construct a critical trajectory emanating from it by taking small steps so that $\vf > 0$ at each step, until we hit another zero. While we do this we can add the line elements $ds = \sqrt{|\phi(z)| } |dz|$ and that would generate the lengths, hence the adjacency matrix $M$, we are looking for.

In passing, we note that the condition of having all non-contractible curves greater than or equal to $2 \pi$ is equivalent the length of each edge of the critical graph to be smaller than $\pi$ for the case of 4-and 5-punctured spheres~\cite{Saadi:1989tb}. This fact has been exploited in previous works by Moeller~\cite{Moeller:2004yy,Moeller:2006cw}. However this condition is not sufficient for higher-punctured spheres, so we opt out to use the generic method described above to make the algorithm manifestly independent of the number of punctures.

\section{Neural networks for accessory parameter and indicator function} \label{sec:ML}

In this section, we describe the neural networks for the accessory parameter $a = a(\xi, \xi^\ast)$ and the indicator function $\Theta(\xi,\xi^\ast)$ in the case of 4-punctured sphere. We show the accessory parameter neural network has successfully learned the analytic behavior for the real moduli described in~\eqref{eq:PlanarAcc} and the symmetry properties described in~\eqref{eq:ConfAcc1}. We emphasize these behaviors haven't been explicitly programmed into our neural network -- they appear as a consequence of the training process. We additionally compare our result for the accessory parameter with the polynomial fit provided by Moeller~\cite{Moeller:2004yy} and observe a good agreement between our results.

Similarly, we test the indicator function neural network by plotting the vertex region $\V_{0,4}$ in the moduli space $\M_{0,4}$ and show our results are consistent with those in the literature~\cite{Belopolsky:1994bj,Moeller:2004yy}. In particular, even though the indicator function neural network outputs values between $0$ and $1$, we observe a sharp transition from Feynman region to vertex region and it is almost always $1$ or $0$ --- as it should be the case for the actual indicator function in~\eqref{eq:IndFunc}.

Lastly, we compute the 4-tachyon contact term in the closed string tachyon potential by performing moduli integration over the vertex region using both trapezoid and Monte-Carlo methods. As mentioned in introduction, we get a good agreement with the results in the literature, providing extra support for our method based on machine learning.

\subsection{Accessory parameter neural network}

\begin{figure}[h]
	\centering
	\includegraphics[width=12cm]{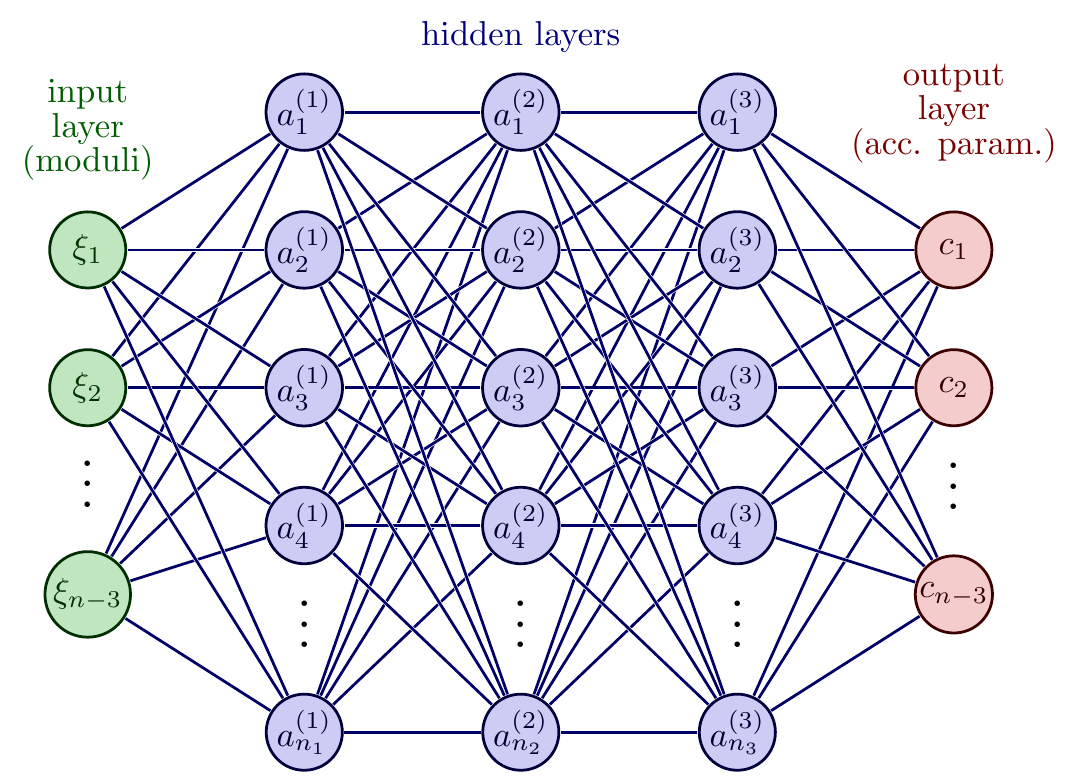}
	\caption{\label{fig:NN}An example of an artificial neural network with 3 hidden layers containing $n_i$ nodes each. It inputs the position of unfixed punctures (moduli) and outputs the (independent) accessory parameters.}
\end{figure}

Artificial neural networks are computing systems inspired by biological neural networks. They consist of number of \textit{layers} and each layer consists of number of \textit{nodes}. A node in a given layer is connected to the nodes in the previous and subsequent layer. An example of a neural network we consider in this work is shown in figure~\ref{fig:NN}.

\begin{figure}[h]
	\centering
	\includegraphics[width=15cm]{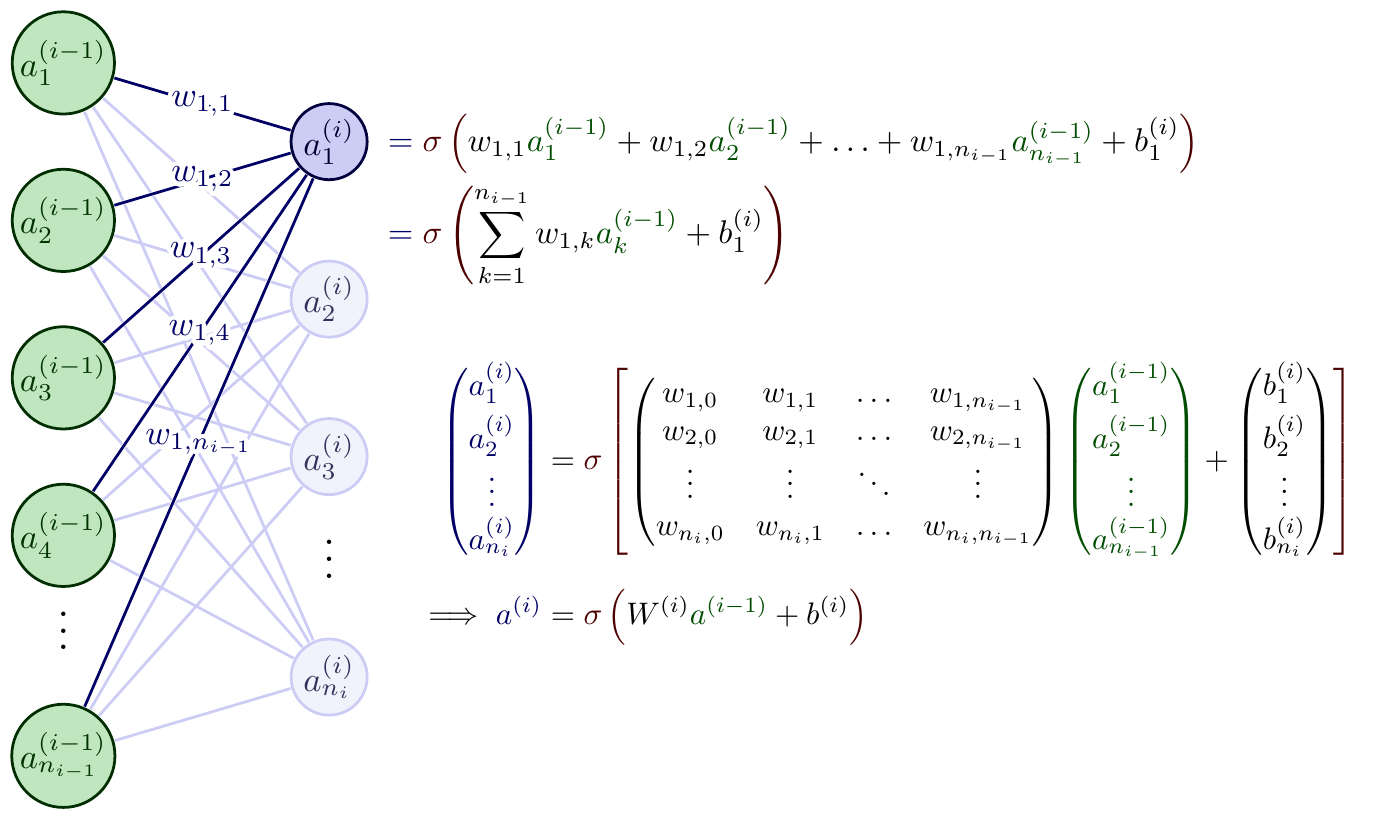}
	\caption{\label{fig:NN_w} The summary of mathematical operations performed by artificial neural networks.}
\end{figure}

At each node, based on the input received from the nodes in the previous layer, a mathematical operation is performed. More specifically, if we denote the collection of inputs received from the $(i-1)$-th layer containing $n_{i-1}$ nodes to a node in the $i$-th layer containing $n_{i}$ nodes as a column vector $a^{(i-1)}$ of length $n_{i-1}$, the nodes in the $i$-th layer would perform \textit{non-linear} transformation
\begin{align}
	a^{(i-1)} \to a^{(i)} = \sigma ( W^{(i)} a^{(i-1)} + b^{(i)} ) \, ,
\end{align}
and transmit this to the nodes in the $(i+1)$-th layer. Here $W^{(i)}$ is a $n_{i} \times n_{i-1}$ matrix, $b^{(i)}$ is column vector of length $n_i$, and the function $\sigma$ is some non-linear function called \textit{activation function}. In the operation above the function $\sigma$ acts on column vectors element-wise. The collection of all $W^{(i)}$ and $b^{(i)}$ for all layers is called \textit{weights} and \textit{bias} respectively and we collectively denote them by $\textbf{W}$ and $\textbf{b}$. Figure~\ref{fig:NN_w} summarizes this procedure. It can be shown that artificial neural networks can approximate class of arbitrarily complicated continuous functions~\cite{Cybenko:1989:ApproximationSuperpositionsSigmoidal,Hornik:1989:MultilayerFeedforwardNetworks, Hornik:1991:ApproximationCapabilitiesMultilayer, Leshno:1993:MultilayerFeedforwardNetworks, Csaji:2001:ApproximationArtificialNeural} for which accessory parameters as a function of moduli are expected to belong.

We are interested to approximate the collection of accessory parameters $c_1, \cdots, c_{n-3}$ uniquely specifying the Strebel differentials on $n$-punctured spheres as a function of moduli~$\xi_1, \cdots, \xi_{n-3}$ using neural networks such as the one shown in figure~\ref{fig:NN}. Since this problem is inherently about complex numbers, we take weights and bias to be complex numbers and use complex neural networks~\cite{Trabelsi:2018:DeepComplexNetworks,Scardapane:2018:ComplexvaluedNeuralNetworks, Bassey:2021:SurveyComplexValuedNeural}. For similar reasons, we take the activation function $\sigma$ to be complex exponential linear unit ($\mathbb{C}ELU$), which is defined for $u \in \mathbb{C}$  by
\begin{align}
	\mathbb{C}ELU (u) \equiv ELU( \mathrm{Re} (u) ) + i ELU( \mathrm{Im} (u) ) \, ,
\end{align}
where $ELU$ is the usual exponential linear unit activation function defined for $x \in \mathbb{R}$ as
\begin{align}
	ELU(x) = \begin{cases}
	x &\text{for} \; x>0 \\
	\alpha (\exp(x)  -1 )  &\text{for} \; x \leq 0
	\end{cases}\, .
\end{align}
Here $\alpha$ is a hyperparameter of the network, and the activation function becomes ReLU for $\alpha = 0$. See appendix~\ref{sec:AppB} for more details on the architecture of the network.

We have implicitly fix the positions of $3$ punctures ($\xi_{n-2}, \xi_{n-1}, \xi_n$) using $PSL(2,\mathbb{C})$ transformation already. We pick these fixed punctures to be at $\xi_{n-2}=0$, $\xi_{n-1} =1$, and $\xi_n = \infty$. Moreover, we solved for three accessory parameters in terms of other parameters and the moduli using~\eqref{eq:QuadCond}. We did these out of convenience for numerical calculations and in order to have a unique answer after training. In terms of 4-punctured sphere, this means we can use the parametrization given in~\eqref{eq:quad4}. In this case the network inputs the position of the unfixed punctures (moduli) $\xi$ and outputs the accessory parameter $a = a (\xi, \xi^\ast)$.

\begin{figure}[h]
	\centering
	\includegraphics[height=9cm, width=11cm]{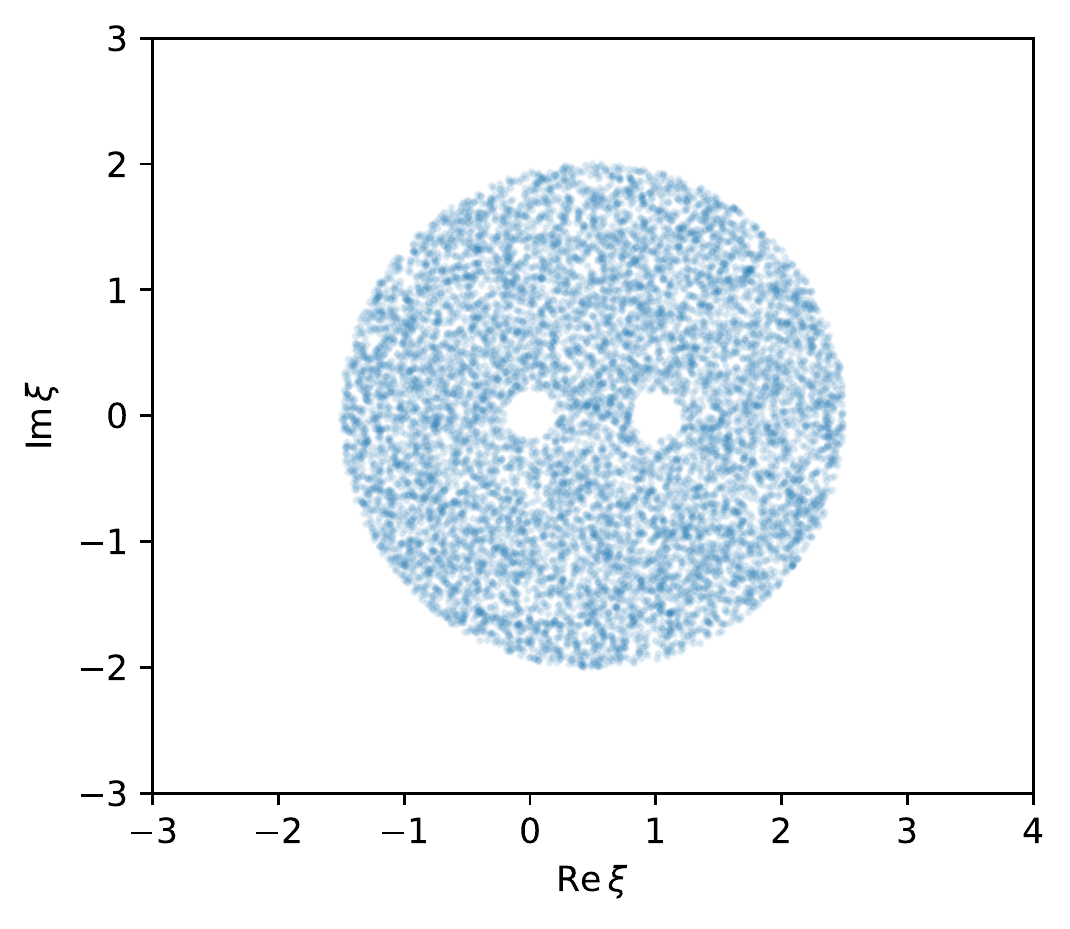}
	\caption{\label{fig:random_data}An example of training set $\mathcal{S}$ for 4-puncture spheres with $|\mathcal{S}| = 10^5$. Notice we have excluded small circles centered at $0,1,\infty$ where 4-punctured sphere is close to degeneration and only sampled points from the remaining triply-connected region (training region).}
\end{figure}

In order to approximate accessory parameters using neural networks we need to adjust the weights $\mathbf{W}$ and bias $\mathbf{b}$ of the network appropriately. Per usual in machine learning, this can be done using iterative (stochastic) gradient descent in the space of weights and bias based on an appropriate averaged loss function. Averaging here is made over finite number of points in the moduli space called \textit{training set} $\cal S$, an example of which is shown in figure~\ref{fig:random_data}. At the end of the gradient descent we end up in some \textit{local} minima of the averaged loss function in the space of weights and bias. The hope is that the resulting network from such local minima would be generic enough to approximate the behavior of accessory parameters, not only for the points in $\cal S$, but \textit{everywhere} on a subset of $\mathcal{M}_{0,n}$ containing the training set $\cal S$. If this is the case, we say the neural network \textit{learned} the accessory parameters.

Before we delve into the specifics of the averaged loss function, let us describe the training set $\mathcal{S}$. For us, the set $\mathcal{S}$ consists of random collection of points uniformly sampled over the moduli space $\mathcal{M}_{0,n}$ excluding the regions where punctures are about to collide. We call this region \textit{training region}. The exclusion condition here is to make sure that the learned behavior for the accessory parameters doesn't get affected by the degeneration behavior of surfaces as our numerical evaluations get unreliable for them.~\footnote{This numerical failure can be understood as follows. As the surface degenerates, some zeros and punctures get closer to each other. This makes evaluating the integral~\eqref{eq:CompLen} numerically unstable and leads to failure in the training.} Recall that we are \textit{not} interested in Strebel differentials on surfaces arbitrarily close to degeneration in the view of CSFT, as it is sufficient to obtain Strebel differentials on the vertex region $\mathcal{V}_{0,n}$ which doesn't contain such surfaces by construction. Hence, as long as we guarantee the training region to cover the vertex region $\mathcal{V}_{0,n}$ and successfully train, we are supposed to be able to get all the geometric data relevant to CSFT.

As mentioned earlier, gradient descent should be performed based on an appropriate averaged loss function. In the case at hand, this is constructed by averaging the function~\eqref{eq:CostFunc} over $\cal S$
\begin{align} \label{eq:Loss}
	L_{0,n}(\mathbf{W}, \mathbf{b}; \mathcal{S}) = {1 \over |\mathcal{S}| }\sum_{\xi \in \mathcal{S}} \mathcal{L}_{0,n} \,  \left( \varphi
	\left( c (\mathbf{W}, \mathbf{b}) , \xi \right) \right) \, .
\end{align}
It is useful to comment on the dependence of indices here. As we have indicated in~\eqref{eq:CostFunc}, the loss function depends on the quadratic differential $\varphi$, which in turn determined by the choice of accessory parameters (collectively denoted as $c$) and moduli (collectively denoted as $\xi$), that is $\varphi = \varphi \left( c, \xi \right)$. But notice that the accessory parameters $c$ are determined by the parameters of the network (weights and bias), hence we have $c = c (\mathbf{W}, \mathbf{b})$. When we average over the points in $\mathcal{S}$, we see the averaged loss function indeed has the dependence shown in~\eqref{eq:Loss}. Note that we have not provided any labels for the points in the training set $\cal S$. So in essence we perform \textit{unsupervised learning} for the accessory parameters using the loss~\eqref{eq:Loss}.

Now we have all the ingredients to train an accessory parameter neural network for 4-punctured spheres. Before we give an example of a training run, let us summarize our strategy to confirm our results. We can test how well the network performs by investigating the loss function $\mathcal{L}_{0,4}$ over the training region. This involves sampling two new sets of points over the training region, called \textit{validation set} if it is used during training or hyperparameter optimization and \textit{test set} for subsequent calculations. If we observe the loss is small not only for the training set, but also over the validation/test sets, we conclude the  network interpolates and declare it has learned the accessory parameter successfully over the training region. We evaluate the loss function to be $8.5 \times 10^{-14}$ for the exact solution for $\xi = e^{i \pi / 3}$ given in~\eqref{eq:exacta}, so we see there is a scale associated with the loss function and its smallness indeed characterizes how close we are to Strebel differential.

\begin{figure}[h]
	\centering
	\includegraphics[width=8.25cm]{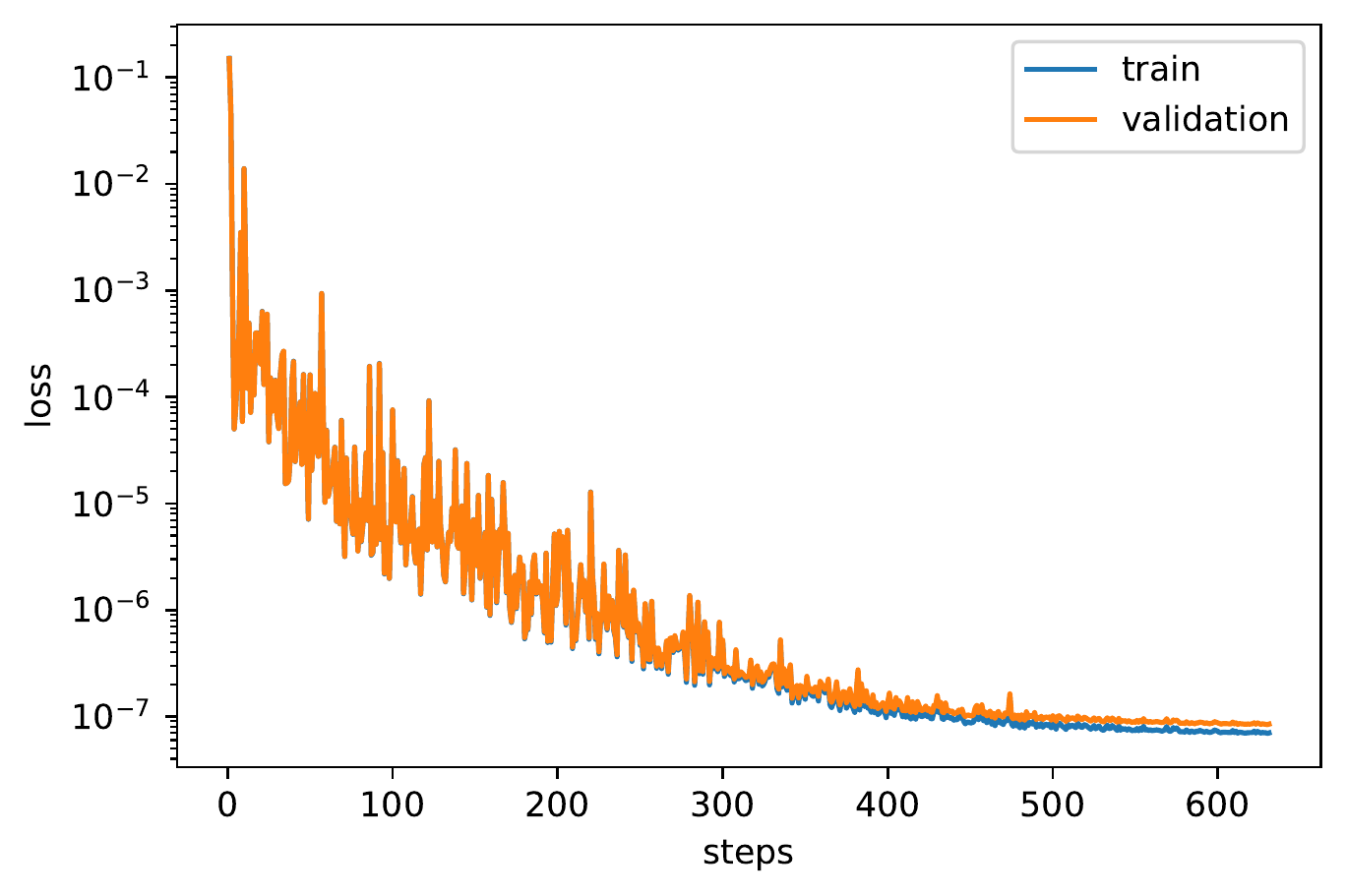}
	\includegraphics[width=8.25cm]{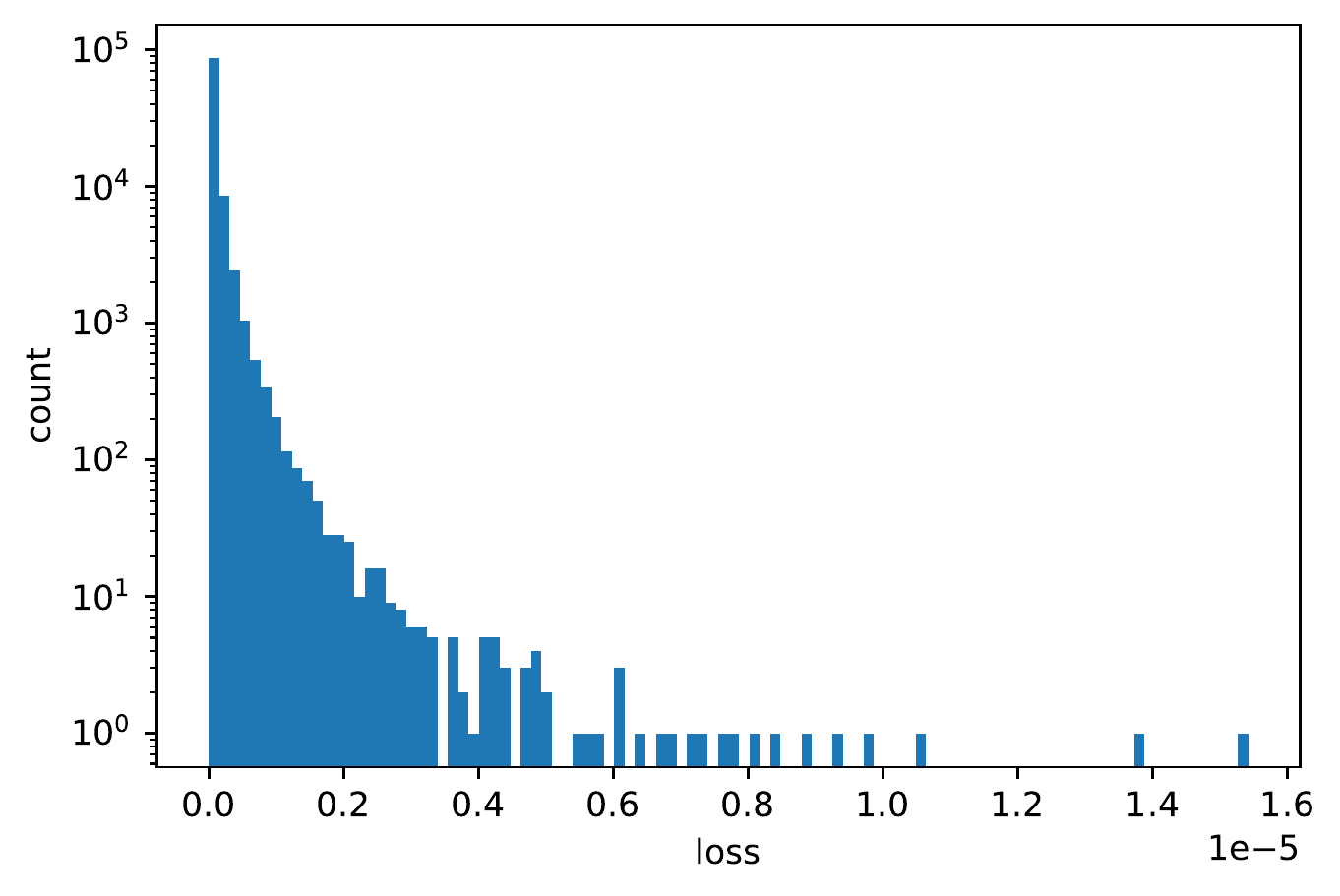}
	\caption{\label{fig:training_c}Training curve (left) and the distribution of loss over the test points in the training region (right) for ``best NN''. As one can observe from the curves on left training was successful and the curve on right informs most points have relatively small losses, except for few outliers.}
\end{figure}

An example of a training based on~\eqref{eq:Loss} is shown in figure~\ref{fig:training_c} and some of its statistics shown in tables~\ref{table:tran_stat} and~\ref{table:tran_stat_1}.~\footnote{
	We characterize the stochastic error of our training procedure after we calculate the off-shell 4-tachyon contact term $v_4$. For clarity of presentation, we just focus on ``best NN'' as a \textit{generic} example. This particular choice is explained at the end of this subsection.}
This particular network has 3 hidden layers with $[512,128,1028]$ nodes each respectively. Training was performed in \texttt{Python} using Google Jax~\cite{jax2018github} by sampling $10^5$ points in the training region shown in figure~\ref{fig:random_data}. We confirm our results with the test set, but also evaluated the loss on the training and validation sets for comparison, and find that they are small and have the same order of magnitude, indicating that the network doesn't overfit and interpolates other points in the training region. Expanded details on the training and the architecture of network can be found in appendix~\ref{sec:AppB}.
\begin{table}[h]
\centering
\begin{tabular}{  | c |  c  | c | c |}
		\hline
		Mean & $8.86 \times 10^{-8}$ & Min & $1.32 \times 10^{-11}$ \\
		\hline
		Median & $3.81 \times 10^{-8}$& Max & $1.54 \times 10^{-5}$ \\
		\hline
\end{tabular}
\caption{The mean, median, minimum, and maximum values for the training loss of the ``best NN''. }
\label{table:tran_stat}
\end{table}
\begin{table}[h]
	\centering
	\begin{tabular}{  | c |  c  | c | c  | c | }
		\hline
		Moduli & Acc. Param. & Loss  & NN Acc. Param. & NN Loss\\
		\hline
		$\xi = {e^{i \pi /3}}$ & $a = 2 + 2/\sqrt{3} i $ & $8.5 \times 10^{-14}$ & $1.9997+1.1548i$ & $6.1 \times 10^{-8}$ \\
		\hline
		$\xi = 1/2$ & $a = 2$ & $7.8 \times 10^{-12}$ & $1.9995-0.0001i$ & $6.6 \times 10^{-7}$  \\
		\hline
		$\xi = -0.2 + 1.5i$ & $ a= 1.12245 + 1.2394i$ & $1.7 \times 10^{-10}$ & $1.12250+1.23965i$ & $7.4 \times 10^{-9}$  \\
		\hline
	\end{tabular}
	\caption{Comparison of the losses and accessory parameters for previously known solutions with the results  ``best NN'' produced. The last point is solved using Newton's Method and is taken from~\cite{Moeller:2004yy}.}
	\label{table:tran_stat_1}
\end{table}

\begin{figure}[p]
	\centering
	\vspace{-1cm}
	\includegraphics[width=10cm]{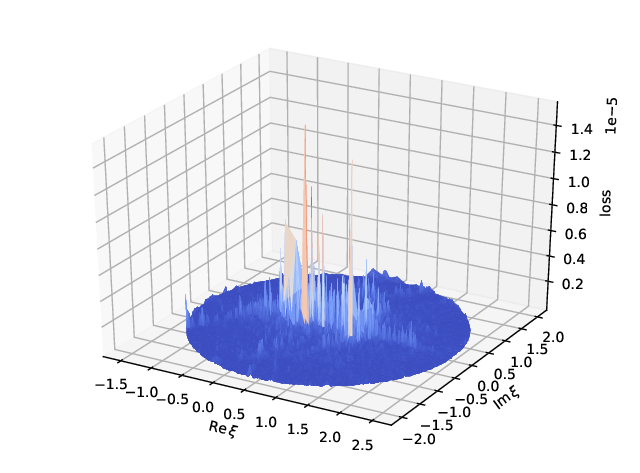}
	\caption{\label{fig:training}The distribution of the loss over the training region. Network underperforms close to the real line.}
	\vspace{1cm}
	\centering
	\includegraphics[width=8cm]{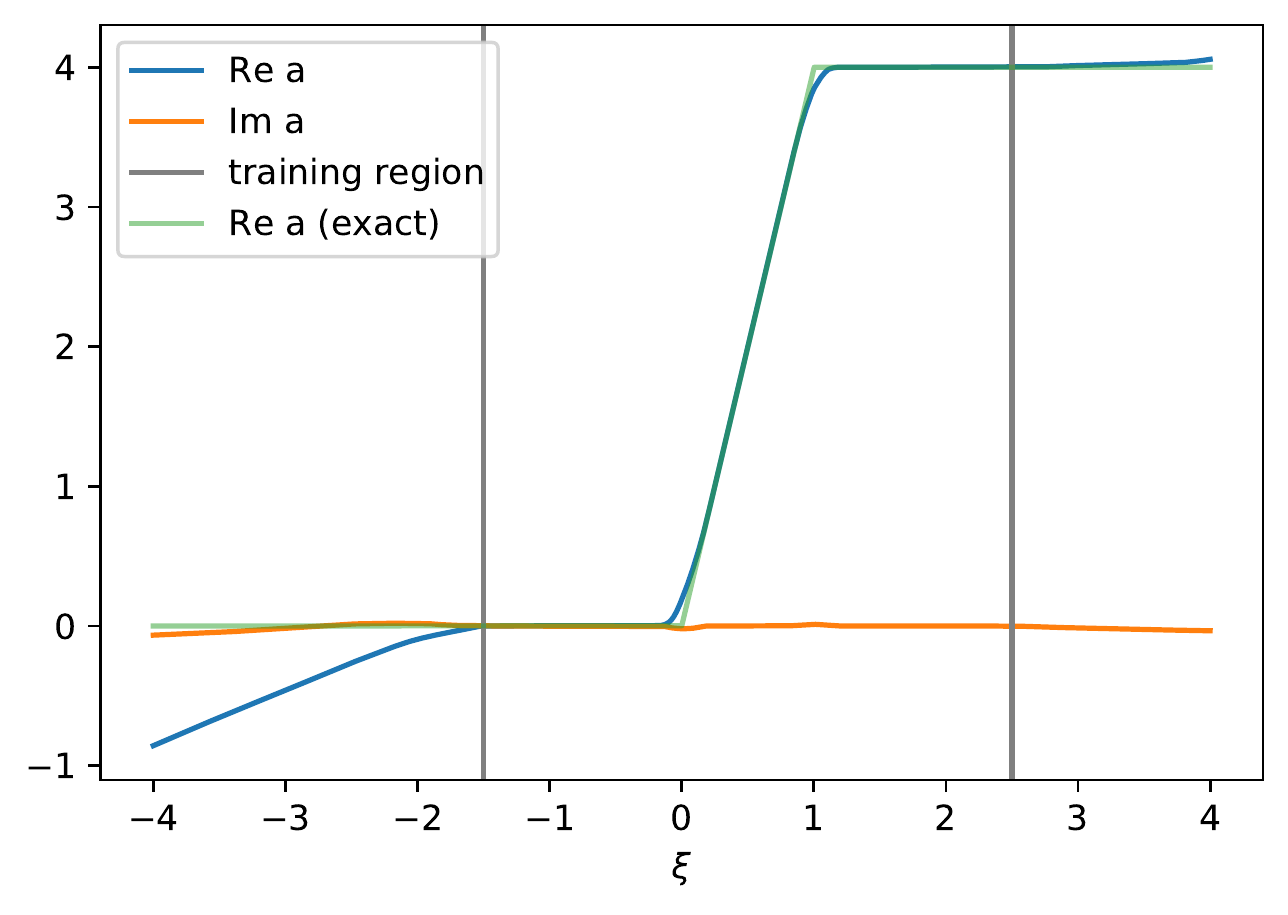}
	\includegraphics[height=5.8cm, width=8cm]{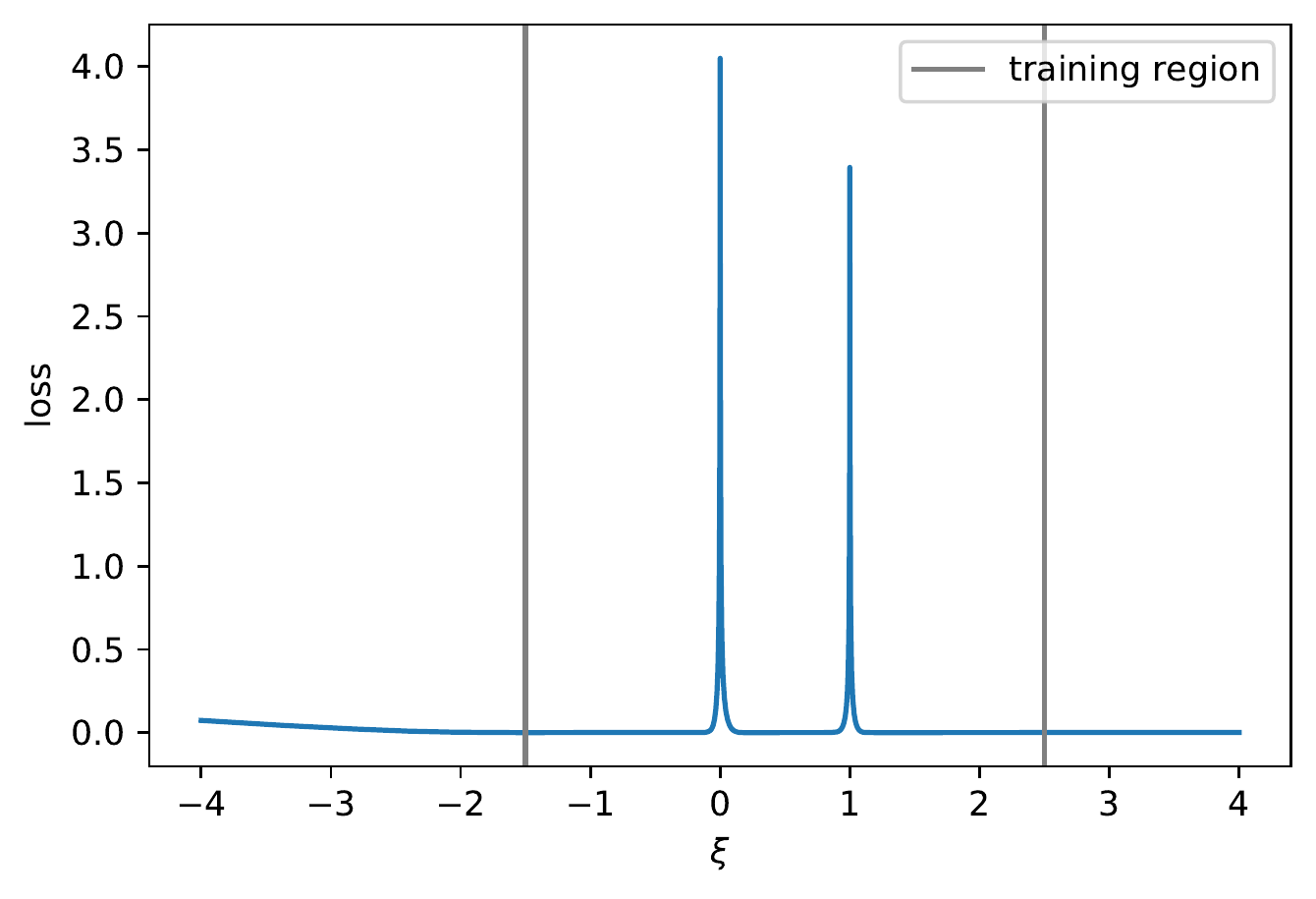}
	\includegraphics[width=10cm]{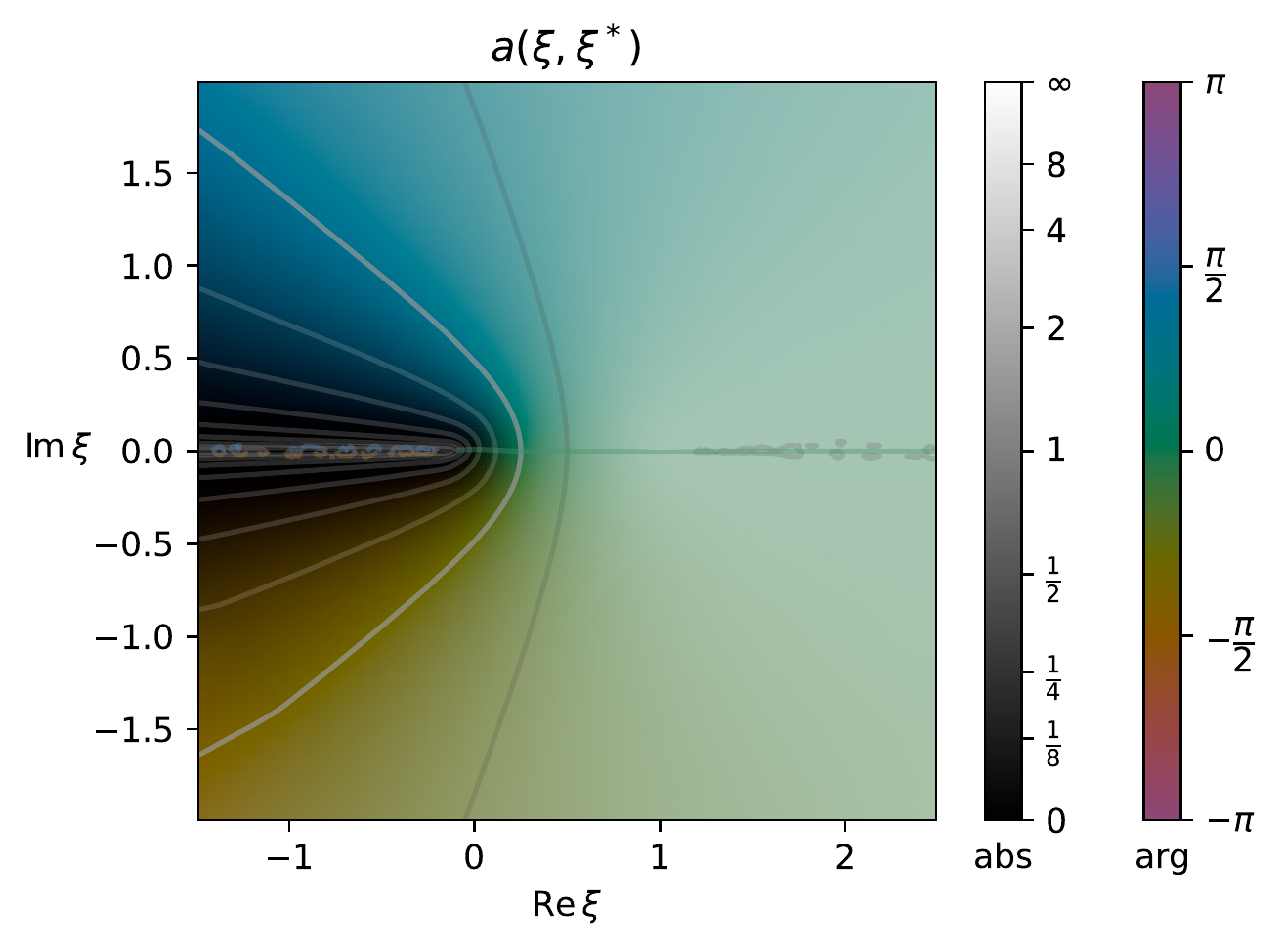}
	\caption{\label{fig:a}The behavior of the accessory parameter compared with the exact solution (top left) and its loss (top right) for the real moduli. Notice the exact and trained behaviors are almost indistinguishable and differ slightly only when the moduli is close to $0$ or $1$ or outside of the training region. Even if this is the case, we see the network was still able to extrapolate away from the training region. The overall behavior of the accessory parameter $a = a(\xi, \xi^\ast)$ is plotted below.}
\end{figure}

Observe from figure~\ref{fig:training_c} that almost all points in the training region have relatively small loss except for few outliers. In fact, by plotting the behavior of the loss function over the training region (figure~\ref{fig:training}) we see these outlier points primarily lie close to the real line --- the region we don't need for CSFT. The reason for this behavior is actually clear: when the moduli is real, some of the terms in the sum given in~\eqref{eq:CostFunc} becomes zero and some of them imposes the equivalent condition as the critical graph degenerates. So relative to other points on the moduli space, the loss function close to the real line is \textit{less} constrained leading to relatively larger loss.
Even if the network relatively underperforms for the real moduli, it correctly generates the analytic behavior described in~\eqref{eq:PlanarAcc}. This, along with the behavior of the accessory parameter over the training region, is shown in~figure~\ref{fig:a}.

\begin{figure}[t]
	\centering
	\includegraphics[width=8cm]{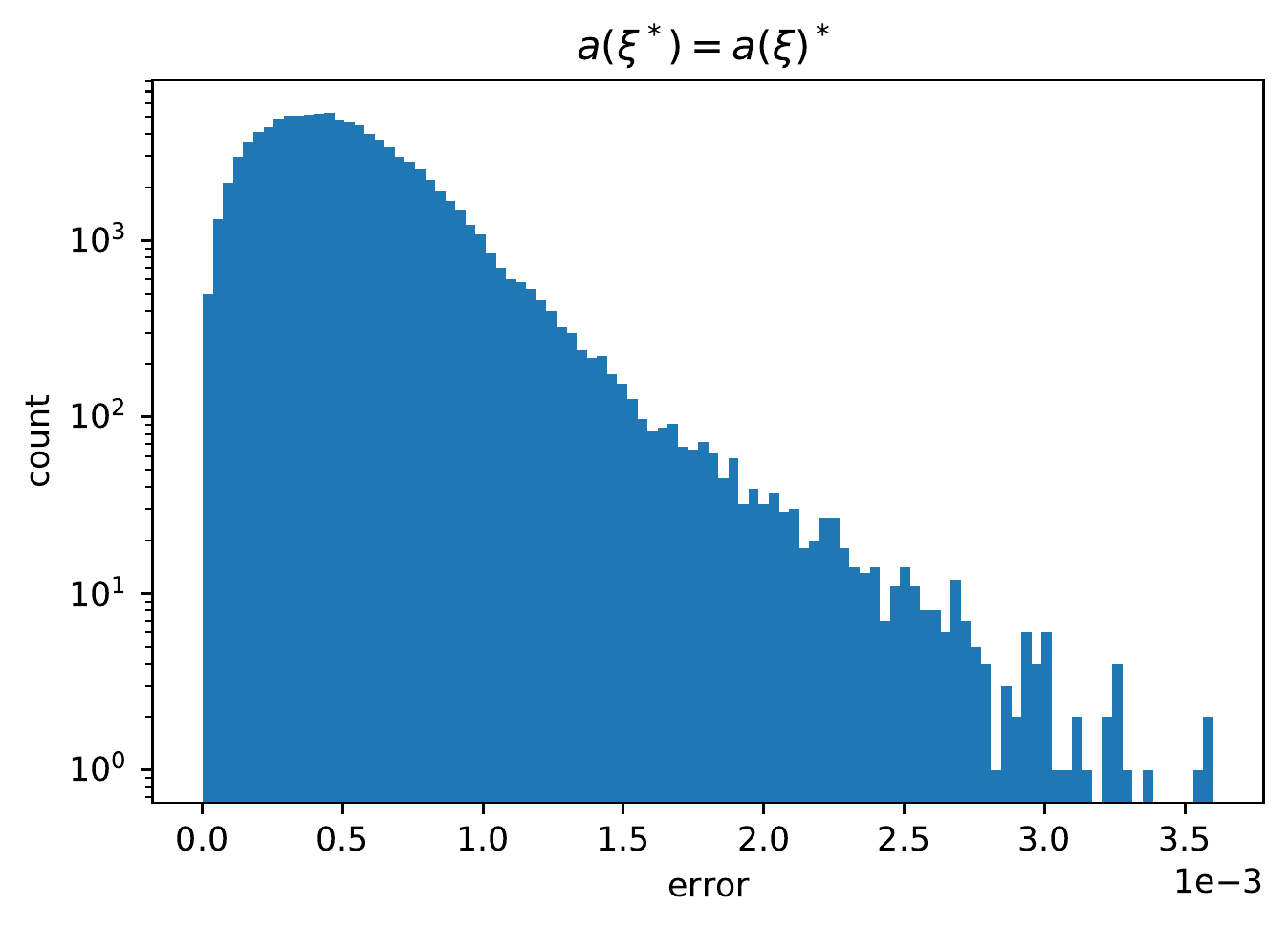}
	\includegraphics[width=8cm]{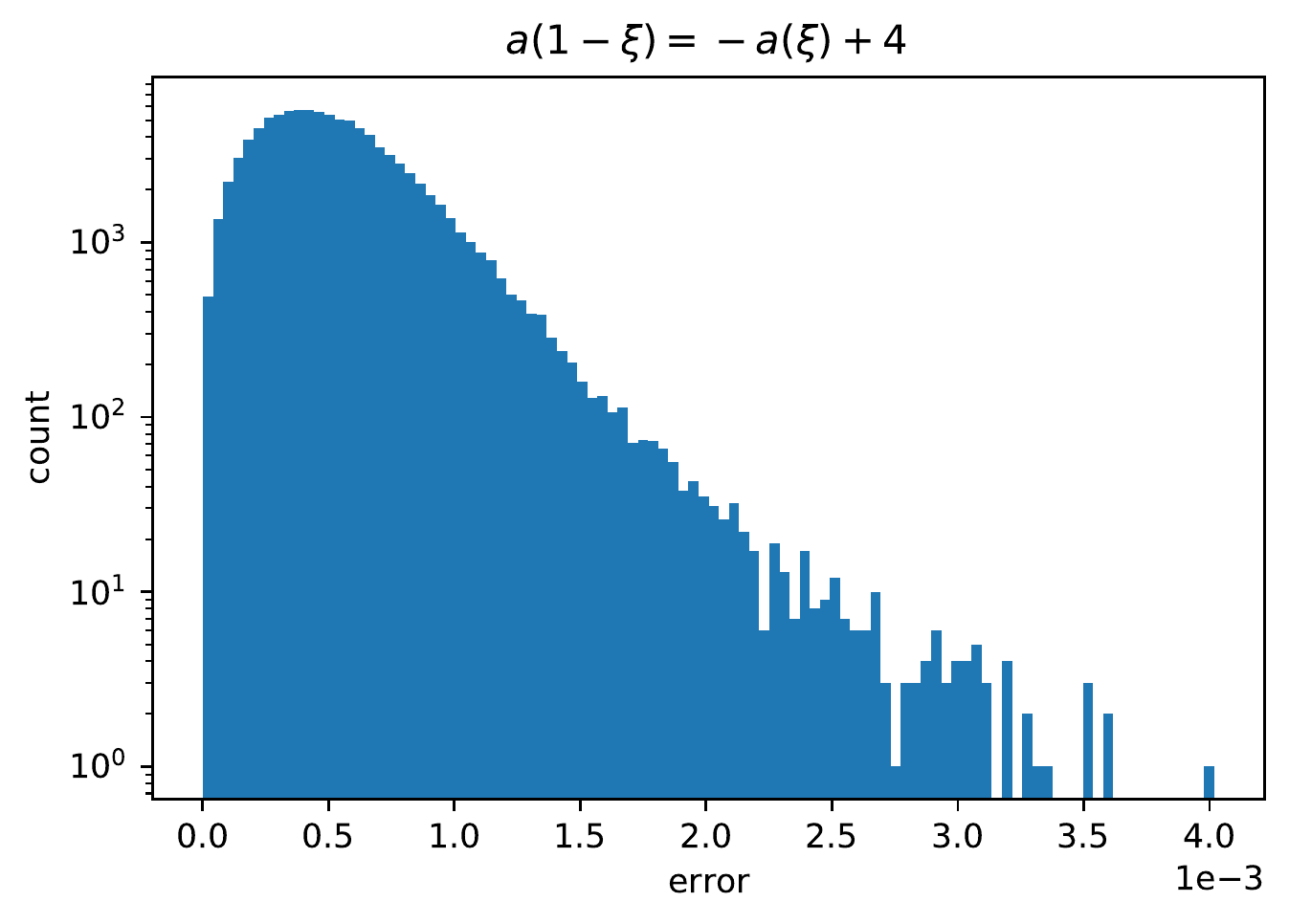}
	\includegraphics[width=8cm]{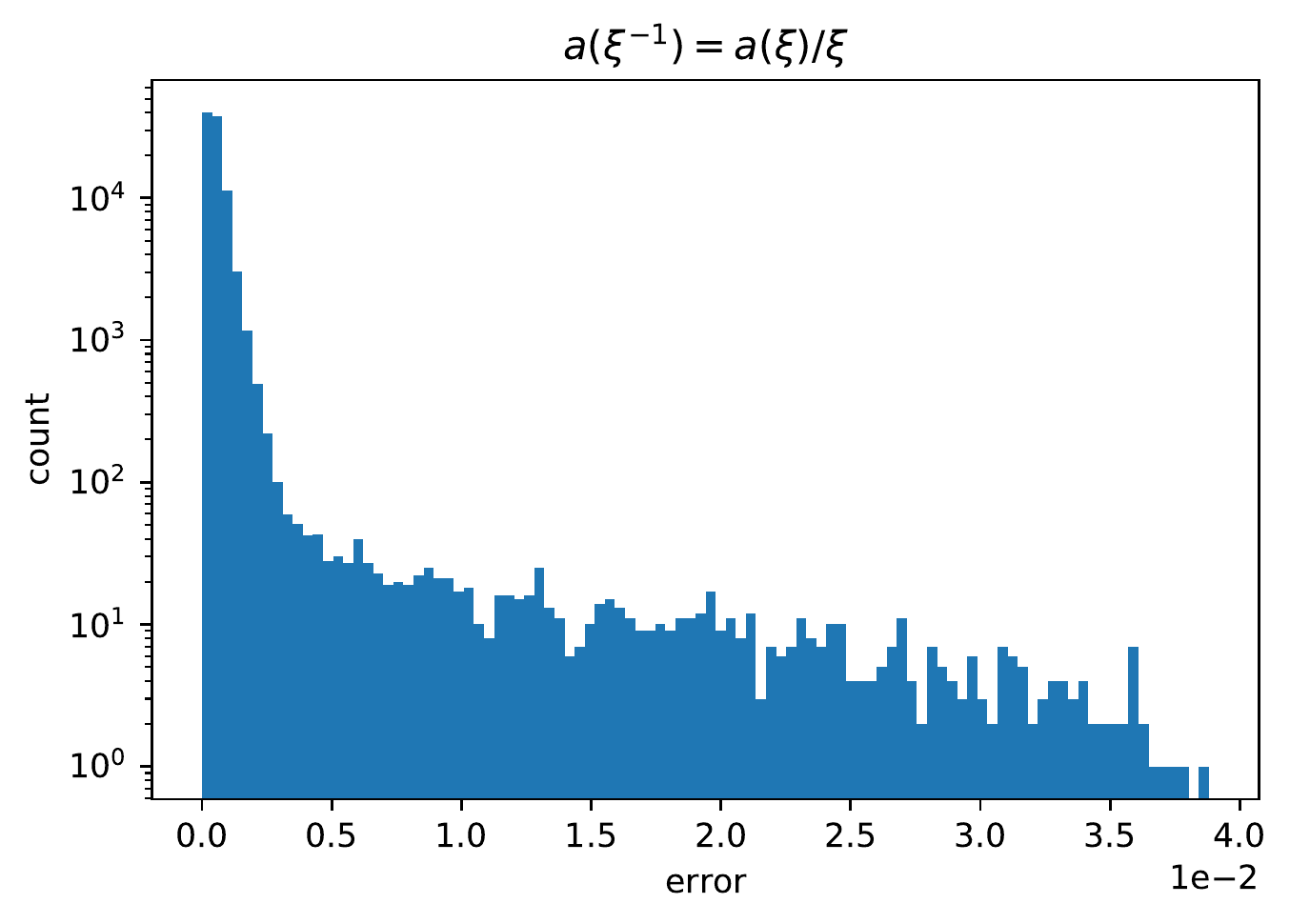}
	\includegraphics[width=8cm]{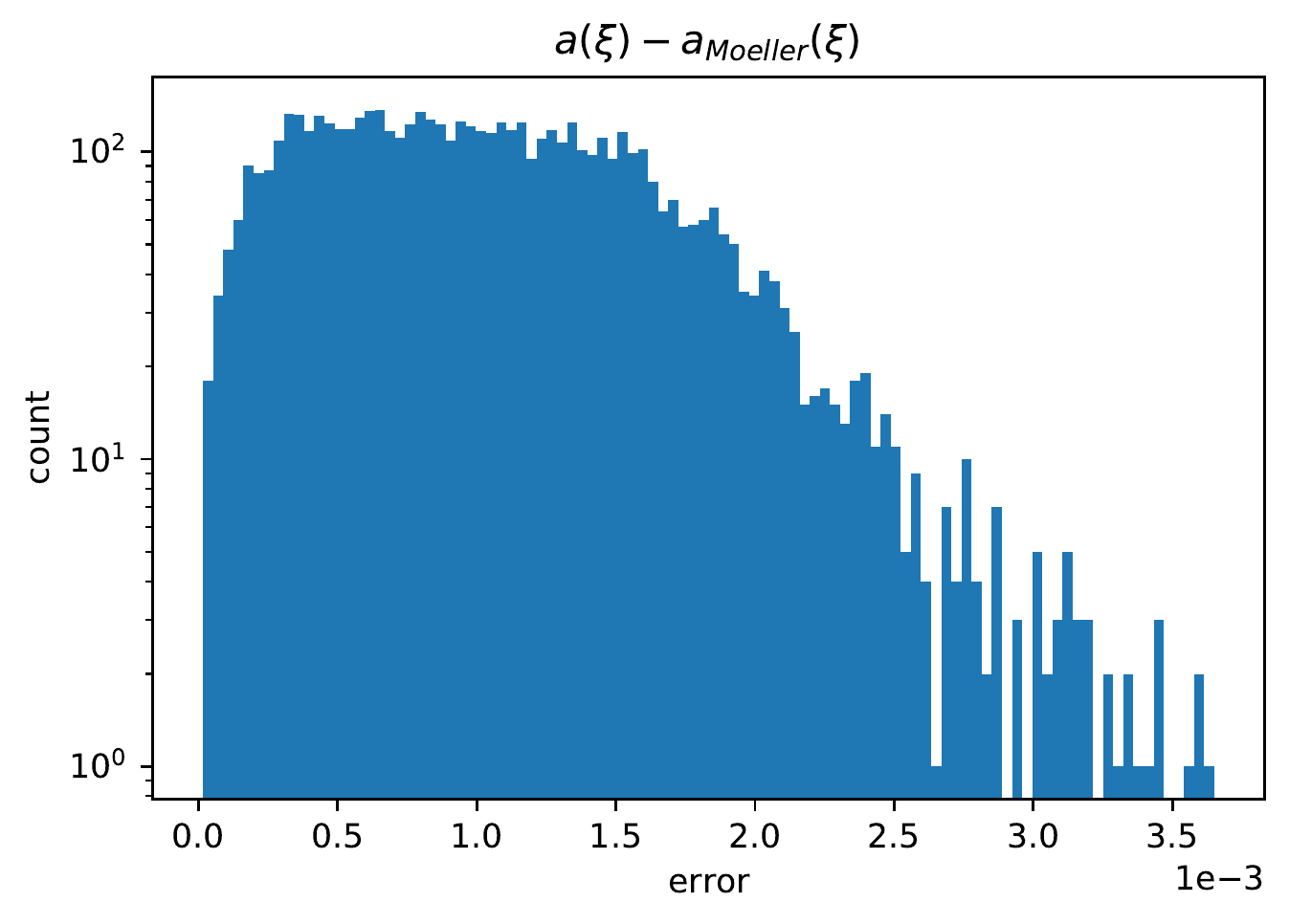}
	\caption{\label{fig:symmetries}Distributions of errors corresponding to involution (top left), shift (top right), and inversion (bottom left) symmetries as well as the comparison of our results with the fit provided by Moeller in~\cite{Moeller:2004yy} (bottom right) over relevant regions.}
\end{figure}

Already from figure~\ref{fig:a}, it is apparent that the network has learned the involution symmetry of $a$. We can quantify this, along with the shift and inversion symmetries given in~\eqref{eq:ConfAcc1} by comparing network's result for a pair of moduli related by symmetry. In order to do that, define the error by
\begin{align}
	\epsilon_{g} (\xi, \xi^\ast)  \equiv |a (g ( \xi ), g ( \xi^\ast ) ) - g(a) (\xi, \xi^\ast)| \, ,
\end{align}
where $g$ represents the symmetry transformations. For example, $g(\xi) = \xi^\ast$ and $g(a) = a^\ast$ for the involution symmetry. Figure~\ref{fig:symmetries} shows the distribution of $\epsilon_{g}$ for points sampled over the training region for three distinct symmetries. As one can see, the errors are quite small and we can conclude the network has learned the symmetries of the accessory parameter \textit{without being explicitly programmed}.

Finally we compare our network's result for the accessory parameter with the polynomial fit provided for $a = a(\xi, \xi^\ast)$ by Moeller~\cite{Moeller:2004yy}. Again, we define the error between our results as
\begin{align}
	\epsilon_{\text{Moeller}} (\xi, \xi^\ast) \equiv |a ( \xi , \xi^\ast  ) - a_{\text{Moeller}} (\xi, \xi^\ast)| \, ,
\end{align}
where $a_{\text{Moeller}}$ is given by the equation (6.9) in~\cite{Moeller:2004yy}. Since the fit in~\cite{Moeller:2004yy} is provided for a subset of the vertex region, we only consider the errors for the points sampled in this subset.~\footnote{We used the indicator function neural network to do this, see next subsection for the details.} Again, we see our results and Moeller's fit are consistent with each other from figure~\ref{fig:symmetries}.

We have listed various evidence for our approach above and they show using machine learning to solve for accessory parameters is sound and the results one gets this way are consistent with the exact solutions as well as with the literature. We have worked with 4-punctured sphere, but we emphasize everything in this subsection admits a trivial generalization to higher-punctured spheres in principle. Thus the results here should be viewed as a \textit{proof of principle}.

Before closing off this subsection, we note that the trained network always interpolates in the training region, but one can ask whether it \textit{extrapolates} to outside of the training region. We observed extrapolation of our networks is not quite as good as their interpolations as it is already somewhat evident from figures~\ref{fig:a}. However we also observed that the better the network extrapolates, the better our results become. So if we would like to specialize to networks among the trained networks, it seems reasonable to us doing this based on how well the network extrapolates for the real moduli and discard the rest of the runs.~\footnote{We quantify this by investigating the behavior for the real moduli between $2 \geq \xi \geq 4$ compared to the solution~\eqref{eq:PlanarAcc} and keep the networks that has order of magnitude smaller average loss. See section~\ref{sec:4tachyon} for more details.} This further defines the ``best NN'' for us: it is the network that extrapolates farthest away on the positive real line and it was the one we chose to report here. In higher-punctured spheres analogous procedure can be repeated by investigating specific degeneration limits.

\subsection{Indicator function neural network}

As we have described in previous subsection, it is possible to obtain accessory parameters as a neural network. Once such representation is in our possession we can solve for the local coordinates and mapping radii as described in section~\ref{sec:Rev}. So all it remains for constructing classical CSFT action is to solve for the explicit description of the vertex region $\V_{0,n}$ over which the moduli integration has to be performed. In this subsection, we train a neural network for the indicator function for the vertex region $\V_{0,4}$, which has already been defined in~\eqref{eq:IndFunc}. This provides an explicit characterization after~\eqref{eq:IndFuncC}. Again, we emphasize that the methods here can be trivially extended to the situation in higher-punctured spheres.

We train the indicator function neural network by performing \textit{supervised learning}. In order to do that, we begin by uniformly sampling points over the training region. However, unlike before, we label these points based on whether they are in the vertex region or not. Remember, a point in the moduli space is in the vertex region if and only if all non-contractible curves in its associated $\vf$-metric has length greater than or equal $2 \pi$. Since the accessory parameter is known, we can compute these lengths using the method described in section~\ref{sec:Rev} and use them to label points: $1$ if all such lengths are greater than or equal to $2\pi$ and $0$ otherwise.

Randomly sampled points in the training region, together with their labels, would form the training set $\mathcal{S}' $ is given by
\begin{align}
	\mathcal{S}' = \big\{ (\xi, \Theta^{(true)}) \,  | \, \xi \in \mathcal{M}_{0,n}, \Theta^{(true)}\in \{0,1\} \big\} \, .
\end{align}
Now the problem of solving for the indicator function~\eqref{eq:IndFunc1} transforms into a binary classification problem. In the view of this, let us call the indicator function neural network $\Theta^{(NN)}_{0,n}$. It inputs the moduli and outputs some value \textit{between} 0 and 1, i.e. $\Theta^{(NN)}_{0,n} : \M_{0,n} \to [0,1]$. So strictly speaking $\Theta^{(NN)}_{0,n}$ is a probability distribution for a given point in $\M_{0,n} $ to be element of the vertex region $\V_{0,n}$. In any case, we are going to observe transition from $0$ to $1$ is relatively sharp when $n=4$. We comment more on this below.

The network $\Theta^{(NN)}_{0,n}$ can be trained to learn the indicator function for $\V_{0,n}$ after performing gradient descent in the space of its weights and bias using the \textit{cross-entropy loss}
\begin{align} \label{eq:CrossEntropy}
L_{0,n}' &= {-1 \over |\mathcal{S}' |} \sum_{ \xi \in \mathcal{S}'}
\bigg[ \Theta^{(true)} (\xi, \xi^\ast) \log
\Theta^{(NN)}_{0,n} (\xi, \xi^\ast)
+ (1-\Theta^{(true)} (\xi, \xi^\ast)) \log (1-\Theta^{(NN)}_{0,n} (\xi, \xi^\ast))
\bigg] \, .
\end{align}
Like in the accessory parameter neural network, we are going to focus only on  four-punctured spheres as a testing ground.

\begin{figure}[t]
	\centering
	\includegraphics[width=8.25cm]{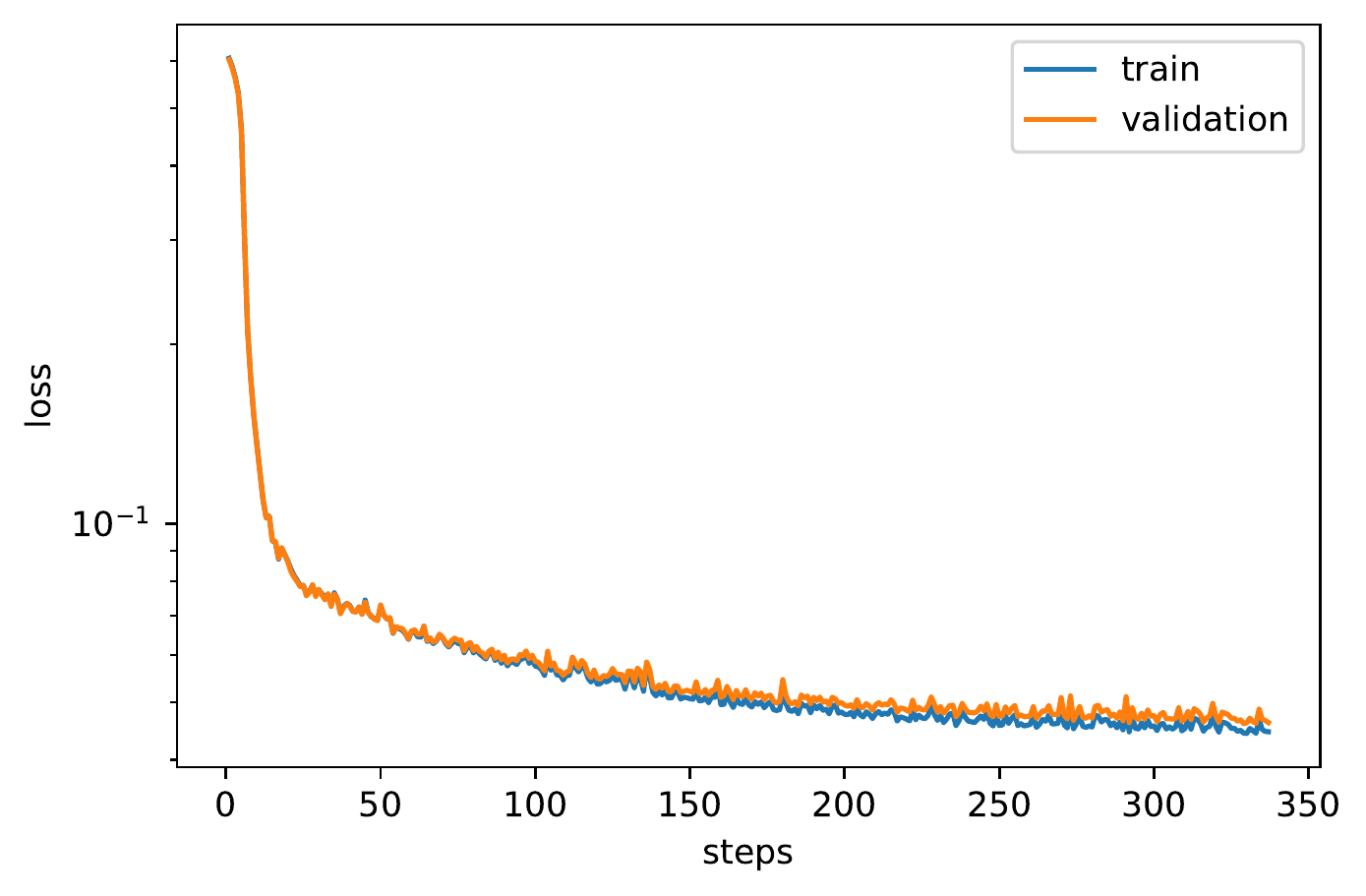}
	\includegraphics[width=8.25cm]{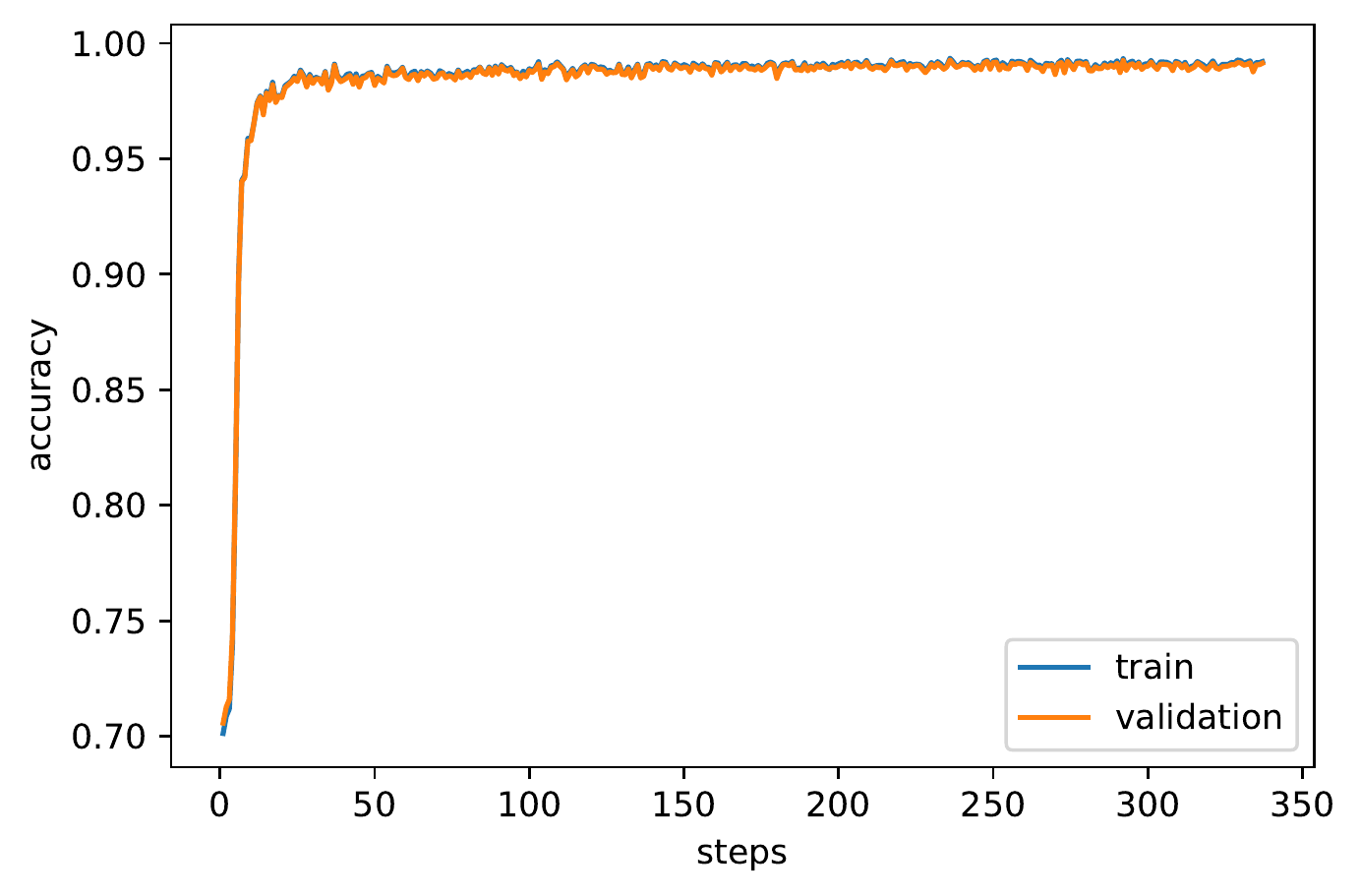}
	\caption{\label{fig:vtrain}The training curve (left) and the progression of accuracy during training (right). High degree of accuracy is achieved for both training and validation sets.}
\end{figure}

\begin{figure}[t]
	\centering
	\hspace{-0.5cm}
	\includegraphics[height=6.5cm, width=9cm]{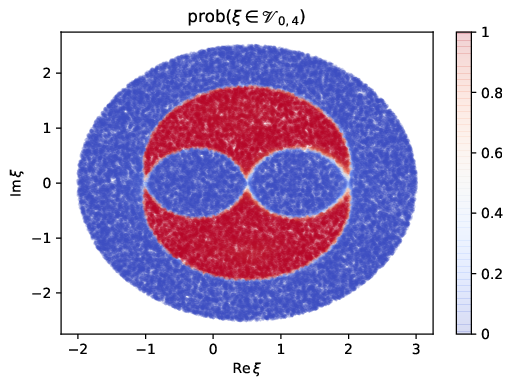}
	\includegraphics[height=6.5cm, width=7.5cm]{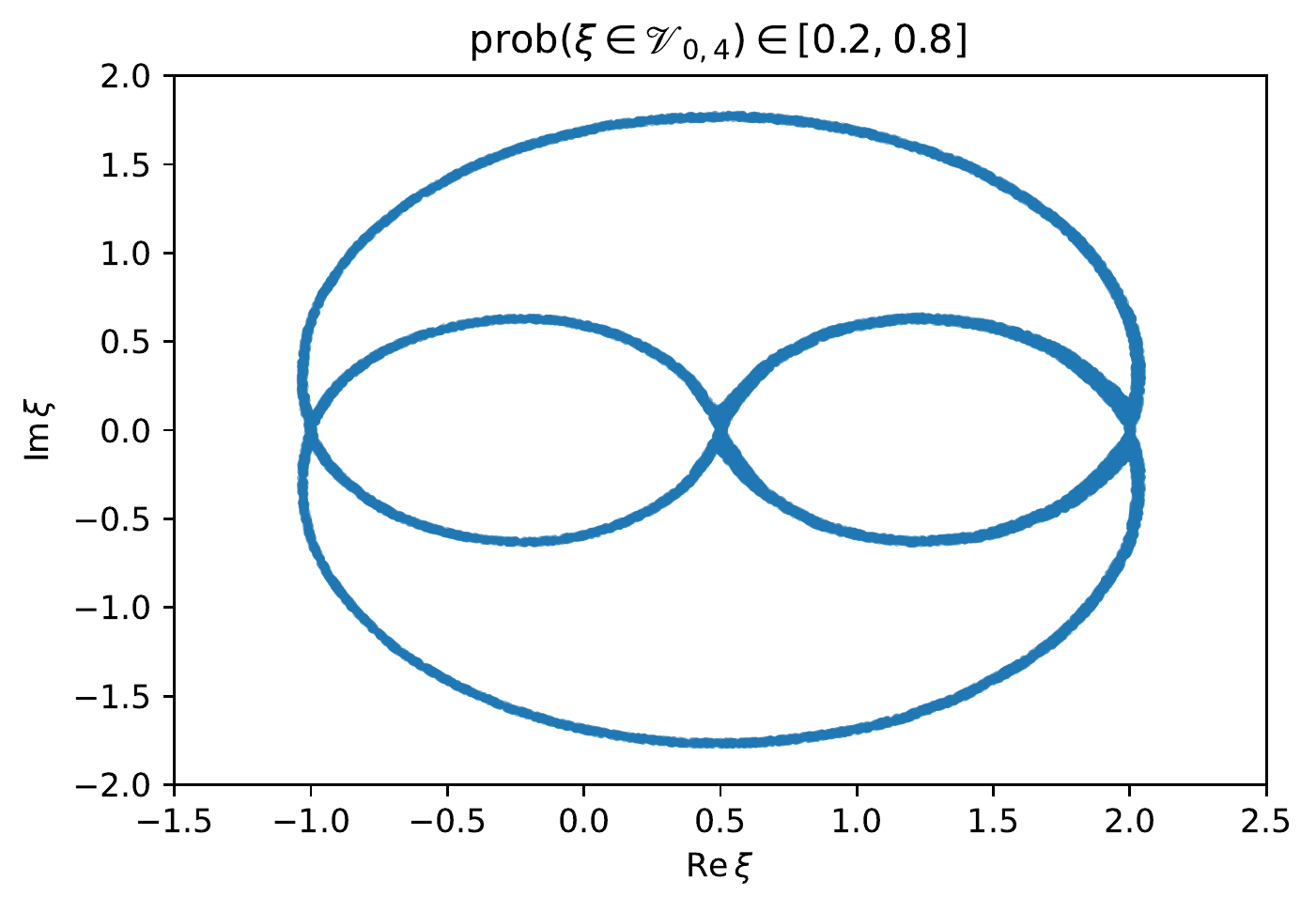}
	\caption{\label{fig:spiderman}The probabilities obtained from the neural network $\Theta^{(NN)}_{0,4}$. }
\end{figure}

The training curve, along with the progression of accuracy during the training, for $\Theta^{(NN)}_{0,4}$ is shown in figure~\ref{fig:vtrain}. For this particular network, we have used the training set $\mathcal{S}' $ constructed using the best NN. Again, the behavior we obtain was generic and such a choice was purely for the presentation purposes. We used $10^5$ points for training. The weights and bias of this networks was chosen to be \textit{real} and we input the complex moduli as a two-dimensional vector. We confirmed our results by checking the loss, as well as accuracy, for both training and validation sets. We have achieved 99.34\% accuracy for the training set, 99.27\% for the validation set, and 99.68\% for the test set.\footnote{It is not common for the test accuracy to be higher than the training and validation accuracy. This may indicate an \textit{underfit}. However the test loss is much higher, see table~\ref{table:stats}.} We observed no overfitting as is evident from figure~\ref{fig:spiderman}. These results show the training was successful. Expanded details on the training and the architecture  can be found in appendix~\ref{sec:AppB}.

Figure~\ref{fig:spiderman} shows the probabilities $\Theta^{(NN)}_{0,4}$ produces. Note that the shape of the region $\V_{0,4}$ shown in~\ref{fig:spiderman} is consistent with the literature~\cite{Belopolsky:1994bj,Moeller:2004yy}.  Moreover, we see the transition from $0$ to $1$ is quite sharp: $\Theta^{(NN)}_{0,4}$ provides a good approximation for the indicator function. Since this is the case, we declare a point in $\M_{0,4}$ is an element of $\V_{0,n}$ when the generated probability is greater than or equal to $1/2$. That is, we declare our indicator function to be
\begin{align} \label{eq:heaviside}
	\Theta_{0,n}(\xi, \xi^\ast) \equiv H \left( \Theta^{(NN)}_{0,n} (\xi, \xi^\ast) - {1 \over 2} \right) \, ,
\end{align}
where $H(x)$ is Heaviside step function.

\begin{figure}[t]
	\centering
	\includegraphics[height=6.5cm, width=7.5cm]{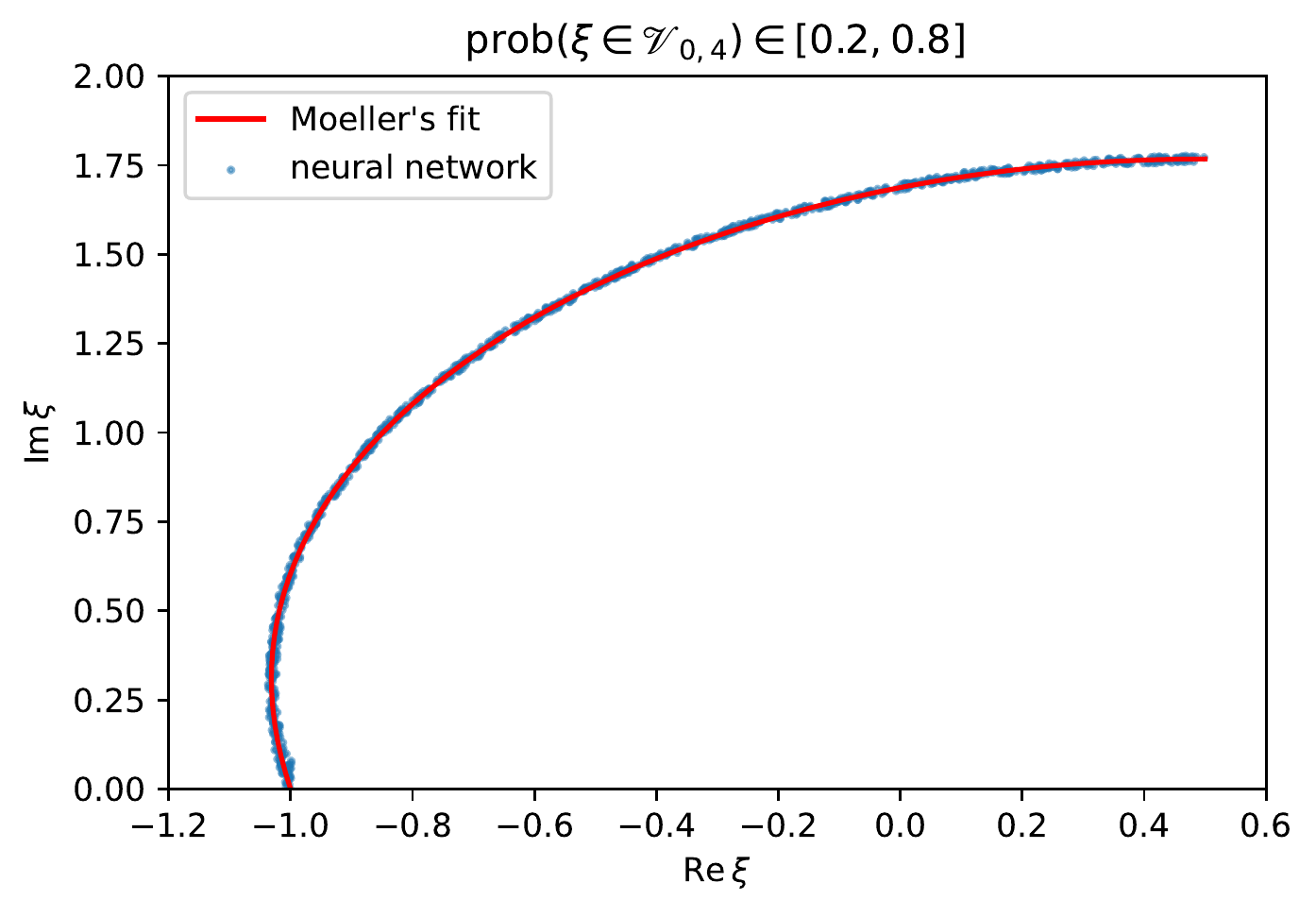}
	\includegraphics[height=6.6cm, width=9cm]{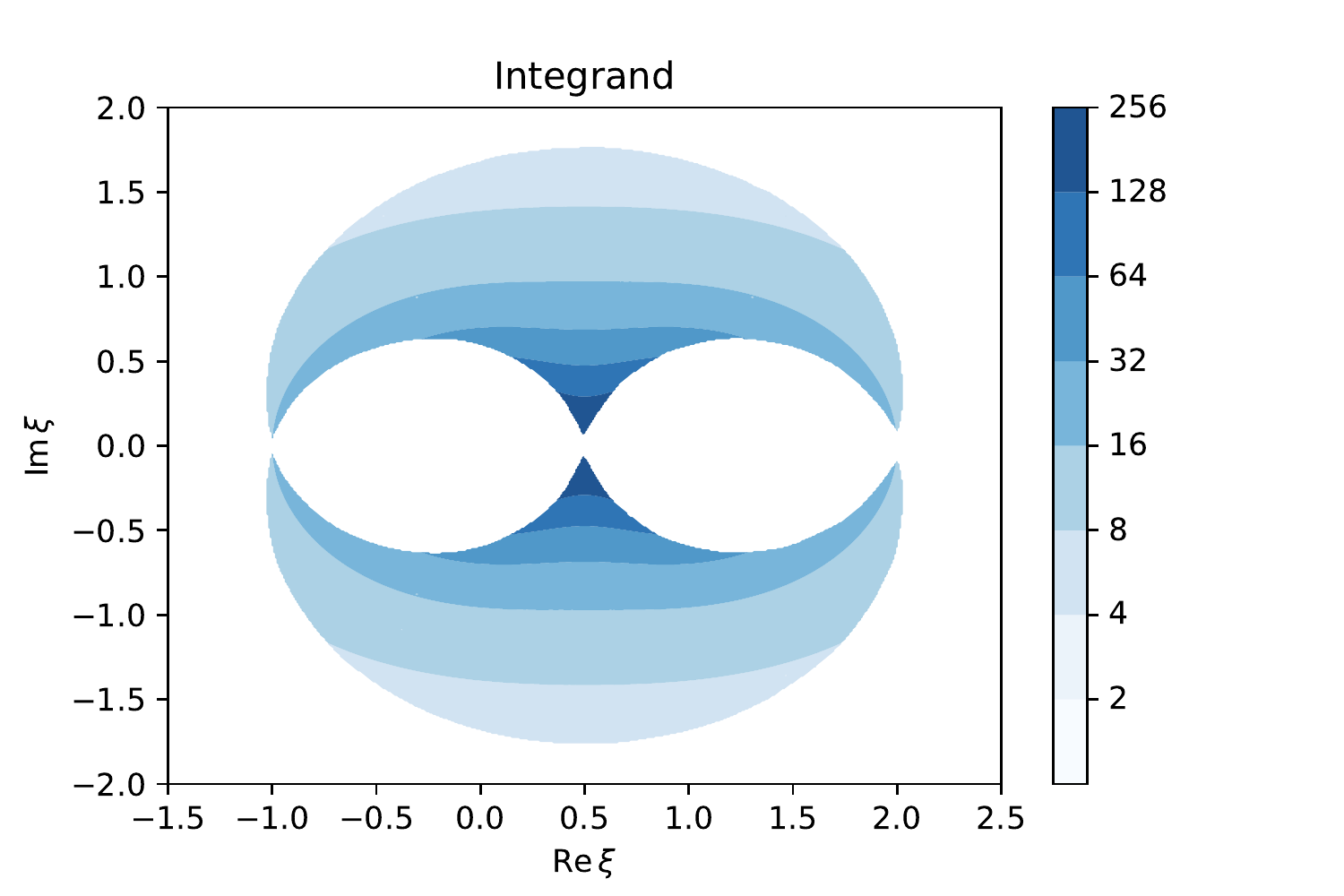}
	\caption{\label{fig:dv_fit}The comparison of $\partial \mathcal{V}_{0,4}$ restricted to $\mathrm{Re} (\xi) \leq 1/2$, $\mathrm{Im} (\xi) \geq 0$, and $|\xi| \geq 1$ between our results and the fit provided by Moeller in~\cite{Moeller:2004yy} (left) and the integrand of the integral~\eqref{eq:contact_term} (right).}
\end{figure}

More quantitatively, we can compare the fit provided by Moeller for the boundary of the vertex region $\partial \mathcal{V}_{0,4}$ restricted to $\mathrm{Re} (\xi) \leq 1/2$, $\mathrm{Im} (\xi) \geq 0$, and $|\xi| \geq 1$ (equation (6.5) in~\cite{Moeller:2004yy}) with the corresponding curve we obtain, i.e. $\Theta^{(NN)}_{0,4} (\xi, \xi^\ast) \approx 1/2$ restricted in similar way. This is shown in figure~\ref{fig:dv_fit}. Again we see a perfect agreement between our results.

\subsection{Off-shell 4-tachyon contact term} \label{sec:4tachyon}

With an explicit description for the vertex region in terms of~\eqref{eq:heaviside}, we now have all the ingredients to find terms in the classical closed string tachyon potential. This is given by~\cite{Belopolsky:1994bj, Moeller:2004yy}
\begin{align}
V(t, \cdots)  = -{1 \over 2} t^2 + \sum_{n=3}^{\infty} {v_n} {t^n \over n!} + \cdots\, .
\end{align}
Here dots represents the terms involving fields other than the tachyon $t$. We are only interested in $v_4$ whose expression can be read from
\begin{align} \label{eq:contact_term}
v_n = (-1)^n {2 \over \pi^{n-3}}  \int_{\mathcal{M}_{0,n} } d^2 \xi \; \Theta_{0,n}(\xi, \xi^\ast) \prod_{i=1}^n {1 \over \rho_i^2} \, ,
\end{align}
where $\Theta_{0,n}$ is the indicator function for $\V_{0,n}$ and $\rho_i$'s are the mapping radii associated with punctures. The convention for the measure is $d^2 \xi = d (\mathrm{Re}\xi) d (\mathrm{Im}\xi)$. Derivation of~\eqref{eq:contact_term} can be found in~\cite{Belopolsky:1994bj}. Note that everything in the integrand of~\eqref{eq:contact_term} can be expressed in terms of two neural networks of previous subsections, so all we need to do at this point is to perform this integration over the moduli space. The integrand of\eqref{eq:contact_term} for $n=4$ is shown in figure~\eqref{fig:dv_fit}. Our results are already presented in table~\eqref{tab:v4}.

The integral is first computed using the trapezoid method along the imaginary and real dimensions, with a grid of $700 \times 700$ points for $\mathrm{Re}(\xi) \in [-1.1, 2.1]$ and $\mathrm{Im}(\xi) \in [-2.1, 2.1]$.  In order to assess the stochastic deviations, we have run the full pipeline (training of the accessory parameter and indicator function networks, and computations of $v_4$) $10$ times. As we mentioned earlier, we observed that networks which perform best are those which generalize well outside the training region on the real axis (as in figure~\ref{fig:a}). As a consequence, we kept only the $4$ networks whose mean loss for $\xi \in [2.5, 4]$ is below $0.1$. This allows determining how $v_4$ varies when we change the random seed, which consists of using different training points, network initializations, and stochastic gradient descent. From the scale of uncertainty of our result for this ($v_4 = 72.320 \pm 0.146$) we see our algorithm is sufficiently stable and produced results consistent with the literature. For the best NN we report $v_4 = 72.396$ using trapezoid method.

For the best NN, we additionally perform a Monte Carlo integration with $2 \times 10^{6}$ points in the vertex region. We report its mean and standard deviation to be $v_4 = 72.366 \pm 0.096$ by evaluating it 5 times. We stress that we use the sharpened indicator function~\eqref{eq:heaviside} in both methods.
Of course, trapezoid method provides a deterministic result for $v_4$ so one may question the point of using Monte-Carlo integration here. However, we note that the moduli integration would be higher dimensional for higher-punctured spheres for which Monte-Carlo integration would be superior to any deterministic method. Here we would like to imitate that case and see how large the resulting errors was due to integration, while still having a baseline for the expected result. We observe our result for $v_4$ is still precise sufficiently even if we use Monte-Carlo integration. The convergence of this integral can be further improved using more points or employing importance sampling.

\section{Discussion and future directions} \label{sec:Con}

In this paper we have characterized 4-string string contact interaction using machine learning by constructing neural networks for the accessory parameter for Strebel differentials and the indicator function for the vertex region in the moduli space. Doing so allowed us to construct the local coordinates associated with the 4-string contact interaction. We tested our pipeline by computing the off-shell 4-tachyon contact term in the tachyon potential. We obtained a good agreement with the results in the literature.

We would like to emphasize few advantages of using machine learning over traditional numerical methods to characterize the local coordinates in CSFT:
\begin{enumerate}
	\item The algorithm presented here is manifestly independent of the number of punctures. So it is in principle possible to repeat the similar construction for $n$-punctured spheres.

	\item Having a neural network representation for the accessory parameters would help to explore properties of Strebel differentials and uncover hidden patterns as neural networks are just approximations to the actual function.

	\item Building string interactions using machine learning would eventually simplify the technical aspects of CSFT calculations, as all the geometric data needed for this is encoded in the neural networks. The neural network weights and architectures will be made public in the future.~\footnote{See \url{https://github.com/melsophos/pysft-nn}. The code will be made available as a package together with~\cite{upcoming_work}} Of course, providing fits for the relevant functions, like Moeller did in~\cite{Moeller:2004yy}, achieves the same goal. However, even in the case of 5-punctured sphere, there is no polynomial fits for all of these functions~\cite{Moeller:2006cw}, and even if we are able to construct one, it is a general wisdom that neural network representation of functions are superior to fits.
\end{enumerate}

We want to briefly comment on the precision of our results. Even though we get quite close to the results in the literature~\cite{Belopolsky:1994bj, Moeller:2004yy}, they differ in the third significant digit and there is still room for improvement. We think this is primarily due to the training for the accessory parameter being not sufficiently precise. Even though we have reached quite low losses during the training, as is evident in figure~\ref{fig:training_c}, it is not as low as one would get using Newton's method. This indicates the training precision has to be improved, at least until we reach the same order for the loss as Newton's method~\cite{Moeller:2004yy}. Indeed, the accessory parameters are used in rest of the computations so it is crucial to minimize its error as much as possible.

This type of problem is unconventional from the perspective of machine learning since high precision is rarely needed (although see the recent work~\cite{michaud2022precision}). The tendency is to \textit{decrease} the float precision. Having a loss plateau around $10^{-7}$ in the training shown in figure~\eqref{fig:a} motivates that we need to use double precision \texttt{float64} instead of simple precision \texttt{float32}. Still, this is not sufficient as the usual optimization techniques have not been designed to handle such scenarios. For instance, both the gradients and learning rate are around $10^{-7}$ at the end of the training which implies that weight updates are effectively frozen. Another instance is that the use of regularization (such as $L_2$) can the loss~\eqref{eq:CostFunc} in later epochs. One may try to circumvent these problems by turning off/decaying the regularization and/or using learning rate restart, that is increasing the learning rate to counter-balance the vanishing gradients for later epochs. We have made a preliminary study on using some of these techniques which resulted in smaller loss. We note that the hyperparameter optimization is difficult for these techniques as one needs to train networks over extended periods to find the long-time effects.

In any case, we think such level of precision, at least for the purpose of establishing existence of closed string tachyon vacuum by level/order truncation, won't be needed. Because if the tachyon vacuum happens to be finely-tuned and requires terms in the tachyon potential to be evaluated precisely, the whole procedure for searching the vacuum by  truncating the theory won't work; as it would presumably require all orders of CSFT to be considered.~\footnote{We thank Martin Schnabl for this argument.}

There are numerous natural directions one can take in future. Here we list some of them seem to us of utmost importance:
\begin{enumerate}
	\item An obvious direction is to generalize this approach to higher-order string contact interactions which we are currently working on~\cite{upcoming_work}. As we have emphasized numerous times throughout the paper, there are no conceptual obstruction doing so, as long as the algorithm, especially the training for the accessory parameters, scales favorably.

	Observe that the number of distinct integrals in~\eqref{eq:CostFunc} scales as $\mathcal{O}(n^2)$ and evaluating all of them may slow down the training for higher-punctured spheres. To remedy this problem, consider following modification to the loss function~\footnote{We thank Barton Zwiebach for suggesting this modification.}
	\begin{align} \label{eq:newloss}
		\mathcal{L}_{0,n} (\vf) \to \mathcal{L}_{0,n}  (\vf)  = {1 \over 2n-5} \sum_{(ij)} ( \mathrm{Im} (\ell(z_i,z_j)) )^2 \, .
	\end{align}
	Here sum over $(ij)$ means the following. We first construct an ordered list of zeros and only compute the complex lengths between zeros adjacent to each other in this list. It is easy to argue this new function preserves the properties of~\eqref{eq:CostFunc}. However, compared to~\eqref{eq:CostFunc}, we use just enough condition to specify Strebel differential at the function's global minimum while being agnostic of the  shape of the critical graph. The number of integrals in~\eqref{eq:newloss} scales as $\mathcal{O}(n)$ and this may speed up the training, possibly at the expense of precision due to imposing less condition on the differential.

	Scaling the algorithm for higher-punctured spheres may also require using more advanced architectures, for example, by including equivariant layers for the unitary group $U(n)$ and complex translations $\mathbb C^n$~\cite{Cohen:2016:GroupEquivariantConvolutional,Cohen:2018:GeneralTheoryEquivariant,Gerken:2021:GeometricDeepLearning}. Furthermore we may also want to represent the local coordinates themselves by a new neural network (and more generally, it would be interesting to understand how to compute conformal maps as neural networks) or use graph neural networks~\cite{Zhou:2018:GraphNeuralNetworks} to extract properties from the critical graphs as they become more complicated with increasing number of punctures.

	\item We have trained a neural network for the indicator function distinguishing the vertex region from the Feynman region. It is also possible to train a network that would distinguish distinct type of degeneration of punctured-spheres from each other. For 4-punctured spheres, this means we can train a network that takes different values for $s,t,$-and $u$-type degenerations.

	Such networks allow us to sample points just from the relevant parts of the Feynman region and based on these points it may be possible to train a network for the Zwiebach differentials using the following modified loss function
	\begin{align}
		\mathcal{L}_{Z} (\vf) &=
		\mathrm{Im}(\ell(z_1,z_2))^2 + \mathrm{Im}(\ell(z_3,z_4))^2 \nonumber \\
		 &\hspace{1.5cm} + (\mathrm{Re}(\ell(z_1,z_2))^2 - \pi^2)^2 + (\mathrm{Re}(\ell(z_3,z_4))^2 - \pi^2)^2  \, ,
	\end{align}
	 and variations thereof (i.e. $z_2 \leftrightarrow z_3$ and $z_2 \leftrightarrow z_4$). Remember Zwiebach differentials have a disconnected critical graph in Feynman region (that is, zeros $(z_1,z_2)$ and $(z_3,z_4)$ no longer connected to each other by a critical trajectory) and this is reflected by eliminating terms such as $\mathrm{Im}(\ell(z_1,z_3))$ and replacing them with terms such as  $(\mathrm{Re}(\ell(z_1,z_2))^2 - \pi^2)^2 $.

	Assuming such network can be trained, we can find the local coordinates associated with Feynman diagrams as well. This is interesting for few reasons. First, \textit{all} off-shell string amplitudes will be characterized using neural networks. But more interestingly, this gives an alternative way to plumbing fixture to obtain such local coordinates. It might be interesting to cross-compare these two methods.

	\item Since we approximated functions relevant to the geometry of 4-string contact interaction as neural networks, we can try performing \textit{symbolic regression} to get an analytic insight into the nature of these functions. In particular, it may be interesting to search for closed form expressions for the accessory parameter and $\partial \V_{0,4}$ shown in figure~\ref{fig:dv_fit}.

	\item Generalizing the ideas in this paper to the case of higher genera, especially using minimal-area vertices, would possibly take a non-trivial effort: it is known not all minimal-area metrics in higher genera arises from Strebel differentials while our loss function is intrinsically about the latter. Still, it may be possible to exploit convex optimization approach to minimal-area problem~\cite{Headrick:2018dlw, Headrick:2018ncs, Naseer:2019zau} to construct a suitable loss function.

	Better yet, one may try solving a version of accessory parameter problem appearing in the case of hyperbolic string vertices~\cite{Firat:2021ukc} using machine learning to construct quantum CSFT. In this case the loss function ought to impose the solutions of Fuchsian equation to have a real monodromy on each non-contractible cycle of a Riemann surface at its minimum. We expect the natural loss function for this problem to be independent of the number of punctures and genera. This approach may even provide novel insights to Fuchsian uniformization considering these two problems are continuously connected.

\end{enumerate}

\section*{Acknowledgments}
We would like to thank Barton Zwiebach for endless discussions and introducing us to the world of quadratic differentials. We also would like to thank Riccardo Finotello, Manki Kim, Fabian Ruehle, and Siddharth Mishra-Sharma for discussions. This material is based upon work supported by the U.S. Department of Energy, Office of Science, Office of High Energy Physics of U.S. Department of Energy under grant Contract Number  DE-SC0012567. This project has received funding from the European Union’s Horizon 2020 research and innovation program under the Marie Sklodowska-Curie grant agreement No 891169.

\appendix
\section{Some details on numerical evaluations} \label{sec:App}

In this appendix we give some details on numerical evaluations. In particular we describe the implementation of continuous square root $\sqrt[\pm]{}$, and the numerical evaluations of the integrals for the complex length~\eqref{eq:CompLen} as well as the mapping radii~\eqref{eq:map_rad}.

\subsection{Continuous square root}

Square root $\sqrt{z}$ has a branch cut on $z$-plane and it usually placed along the negative real axis. However, for numerical evaluations of the integrals, such as the one given in~\eqref{eq:CompLen0}, it is advantageous to define a \textit{continuous square root} (denoted by $\sqrt[\pm]{}$) by taking the domain of the complex square root to be the double cover of $\mathbb{C} \setminus \{0\}$ rather than the complex plane with a branch cut. In this domain $\sqrt[\pm]{}$ is a holomorphic function. This way of evaluating square root is inspired by~\cite{Moeller:2004yy}.

Note that the global sign of $\sqrt[\pm]{}$ is ambigious. That is for $z_1 = z_2$, we have $\sqrt[\pm]{z_1} = \pm \sqrt[\pm]{z_2} $ depending on which sheet of the domain we are evaluating $\sqrt[\pm]{}$. As we have stated in the main text, this only leads to technical issues in our numerical evaluations which we overcome by simply squaring the resulting expressions and/or correcting the global sign a posteriori.

In practice, the continuous square root $\sqrt[\pm]{z}$ is evaluated on a given path $z(t)$ by discretizing it as $\{z_i\}_{i =1}^k$ and comparing the complex square roots $\sqrt{z_i}$ and $\sqrt{z_{i+1}}$. More precisely, one is instructed to compute the difference $ |\sqrt{z_{i+1}} - \sqrt{z_{i}} |$ at each step and choose the sign for $\sqrt[\pm]{z_{i+1}} = \pm \sqrt{z_{i+1}}$ such that the said difference is minimum. This way we compensate any branch crossing for complex square root by flipping its sign and get continuous square root instead. As long as the step size is small relative to the distances between $z_i$ and the branch points, the resulting sequence $\{\sqrt[\pm]{z_i}\}_{i =1}^n$ provides a good approximation to $\sqrt[\pm]{z(t)}$.

\subsection{The integral for the complex length}

The complex length~\eqref{eq:CompLen} in the case of $n$-punctured sphere takes the following form
\begin{align}
	\ell(z_i, z_j)
	= \int_{z_1i}^{z_j} \sqrt[\pm]{\phi(z)}
	= \int_{z_i}^{z_j} { \sqrt[\pm]{- (z-z_1) \times \cdots \times (z-z_{2n-4}) } \over z (z-1) (z-\xi_1) \times \cdots \times (z-\xi_{n-3})} dz \, .
\end{align}
Here $z_1, \cdots z_{2n-4}$ are (not-necessarily distinct) zeros of the quadratic differential determined by the accessory parameters $c_i$ and the moduli $\xi_1, \cdots \xi_{n-3}$. For convenience we have fixed the positions of three punctures to $\{0,1, \infty \}$ and solved their accessory parameters using a version of~\eqref{eq:QuadCond}.

We take the integration path to be the straight line between the zeros $z_i, z_j$~\eqref{eq:IntPath} and this gives
\begin{align} \label{eq:NumInt}
	\ell(z_i, z_j)
	= \left({ z_j -z_i \over 2} \right)^2
	\int_{-1}^{1}  { \sqrt[\pm]{ p_{ij}(z(t)) }\over q(z(t))}  \sqrt{1- t^2} dt  \, .
\end{align}
Above we defined the following polynomials
\begin{align} \label{eq:AppPol}
	p_{ij}(z) \equiv \prod_{\substack{k = 1 \\  k \neq i,j}}^{2n-4} (z-z_k) ,
	\hspace{3em}
	q(z) \equiv \prod_{k=1}^{n-1} (z - \xi_k)
	\, ,
\end{align}
In order to make the discussion uniform we have further defined $\xi_{n-2} \equiv 0$ and $\xi_{n-1} \equiv 1$ above. Note that we have taken out the factors $(z-z_i)$ and $(z-z_j)$ from the product in $p(z)$ and we have used the usual square root for $\sqrt{1-t^2}$ as $1-t^2 \geq 0$ for $-1 \leq t \leq 1$.

The goal now is to numerically evaluate the integral~\eqref{eq:NumInt}. In order to do this accurately we have to subtract possible pole contributions and add their analytic expressions. That is, we have to evaluate
\begin{align} \label{eq:AppInt}
\ell(z_i, z_j)
&= \left({ z_j -z_i \over 2} \right)^2
\left[ \int_{-1}^{1} \left( { \sqrt[\pm]{ p_{ij}(z(t)) }\over q(z(t))}
- \sum_{k=1}^{n-1} {s_k r_k \over z(t) - \xi_k} \right ) \sqrt{1- t^2} dt
\right] \nonumber \\
&\hspace{13em} + \left({ z_j -z_i \over 2} \right)^2  \sum_{k=1}^{n-1} s_k r_k \int_{-1}^{1} {\sqrt{1- t^2} dt  \over z(t) - \xi_k}
\, ,
\end{align}
where
\begin{align}
	r _ k \equiv p_{ij}( \xi_k) \prod_{\substack{j = 1 \\  j \neq k}}^{n-1} {1 \over \xi_k - \xi_j } \, ,
\end{align}
and $s_k$'s are signs that should be chosen so that
\begin{align}
	\max_{-1 \leq t \leq 1} \left[ { \sqrt[\pm]{ p_{ij}(z(t)) }\over q(z(t))}  - \sum_{k=1}^{n-1} {s_k r_k \over z(t) - \xi_k}  \right] \, ,
\end{align}
is minimal. This additional step was necessary: the sign of the integral~\eqref{eq:NumInt} is ambigious due to continuous square root. Notice there are $2^{n-1}$ combinations to check.

Once $s_k$'s are determined the first line of the integral~\eqref{eq:AppInt} can be evaluated by trapezoid method using $201$ grid points. It is possible to have cancellation errors for this integral, as we may need to subtract two large quantities close to the punctures, but we have observed this has not caused any problems during the training of the accessory parameter neural network.

The second line of the integral~\eqref{eq:AppInt} can be analytically evaluated as
\begin{align}
	\int_{-1}^1 {\sqrt{1-t^2} \over z(t)- \xi_k} dt
	&= {2 \over z_j - z_i} \int_{-1}^1 {\sqrt{1-t^2} \over t + u}
	= {2 \pi \over z_j - z_i} ( u - i \sqrt{u-1} \sqrt{u+1}) \, ,
\end{align}
where
\begin{align}
	u \equiv {z_i + z_j - 2 \xi_k \over z_j - z_i} \, .
\end{align}
In these expressions the square root is taken on the principal branch. After doing these one has to take the imaginary part of the complex length, square it, and sum over all pairs of zeros for the loss function~\eqref{eq:CostFunc}.

\subsection{The integral for the mapping radii}

The mapping radii integral~\eqref{eq:map_rad} isn't amenable to numerical evaluations. In order to put it into a better form let us first consider the following identity
\begin{align}
	\lim_{ \epsilon \to 0} \left[
	\log |\epsilon| -
	\left(
	- {1 \over \sqrt{z_i-\xi_k} } \int_{\xi_k + \epsilon}^{z_i} {\sqrt{z_i-z} \over z - \xi_k} dz
	+ 2 (\log 2 - 1) + \log|z_i - \xi_k|
	\right)
	\right] = 0\, .
\end{align}
Here $\epsilon \in \mathbb{C}$ is assumed lying on the straight line from the puncture at $z=\xi_k$ to the zero at $z=z_i$ and the branch of square root is adjusted so that this limit exists. Using this in~\eqref{eq:map_rad} drops out any need for $\epsilon \to 0$ limit as
\begin{align}
	\log \rho_k &= \lim_{\epsilon \to 0} \left[ \mathrm{Im} \int_{\xi_k + \epsilon}^{z_i} \left(
	\sqrt[\pm]{\phi(z)} - {i \over \sqrt{z_i-\xi_k} } {\sqrt{z_i-z} \over z - \xi_k}
	\right) dz \right] + 2 (\log 2 - 1) + \log|z_i - \xi_k|  \nonumber \\
	&= \mathrm{Im} \int_{\xi_k}^{z_i} \left(
	\sqrt[\pm]{\phi(z)} - {i \over \sqrt{z_i-\xi_k} } {\sqrt{z_i-z} \over z - \xi_k}
	\right) dz + 2 (\log 2 - 1) + \log|z_i - \xi_k|
	\, ,
\end{align}
Parametrizating the straight line by $z(t) = z_i + t (\xi_k - z_i)$ for $0 \leq t \leq 1$ and exponentiating we get
\begin{align} \label{eq:AppMap}
	\rho_k = {4 \over e^2} |z_i -\xi_k| \exp\left[
	\mathrm{Im} \left( (z_i -\xi_k) \int_0^1 \left(
	 {\sqrt[\pm]{v_{k,i}(z(t))} \over q(z(t))} - {i \over z(t) - \xi_k} \right) \sqrt{t} dt
	\right)
	\right] \, .
\end{align}
The polynomial $q(z)$ is defined in~\eqref{eq:AppPol} and we further define
\begin{align}
	v_{k,i}(z) \equiv - (\xi_k - z_i) \prod_{\substack{j=1 \\ j \neq i }}^{2n-4} (z-z_j) \, .
\end{align}
With the choice of sign for continuous square root made below~\eqref{eq:InvLocCo} the integral~\eqref{eq:AppMap} is finite. In practice though, we would impose this choice by \textit{demanding} the mapping radii has to be finite as it is more convenient to do numerically.

The integral~\eqref{eq:AppMap} computes the mapping radii associated wit the finite punctures. For the mapping radius associated with the puncture at $z = \infty$, we invert the position of punctures to $1/\xi_k$, adjust the accessory parameters (akin to ~\eqref{eq:ConfAcc1}) and calculate the mapping radius associated with the puncture at the origin. Since the mapping radii associated with the origin is invariant under inversion, we obtain the mapping radii associated with the puncture at $z = \infty$.

\begin{figure}[t]
	\centering
	\includegraphics[width=8.25cm]{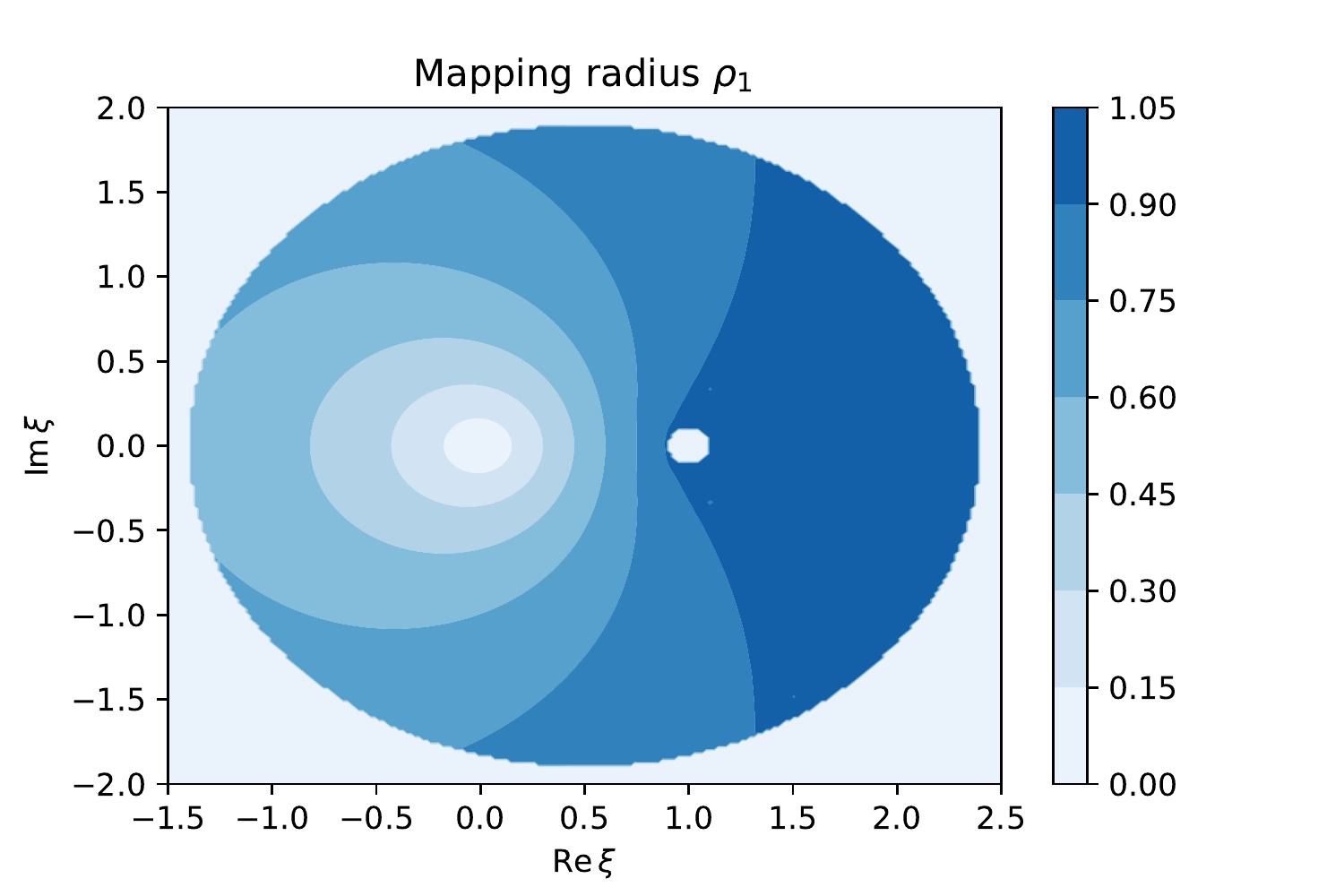}
	\includegraphics[width=8.25cm]{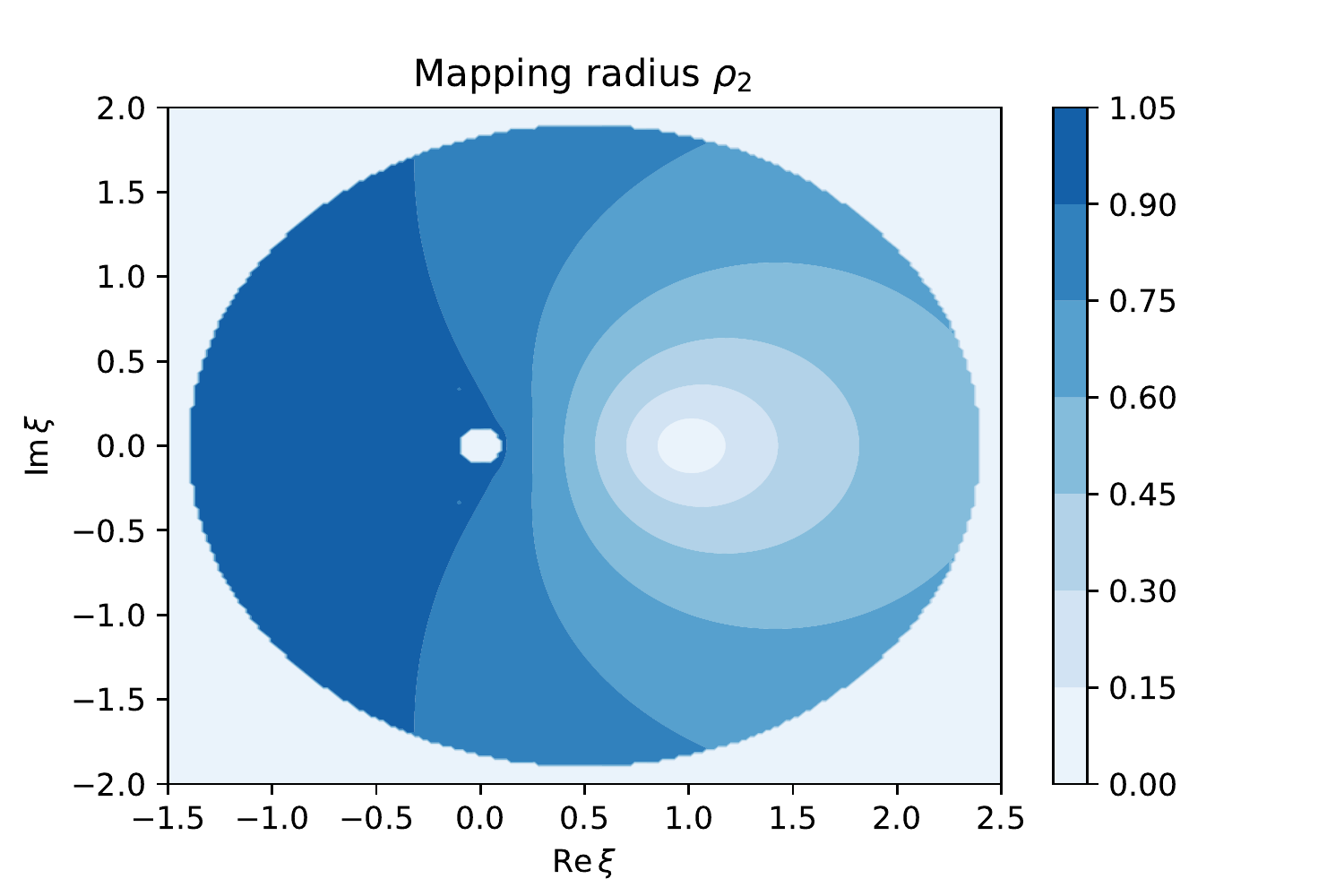}
	\includegraphics[width=8.25cm]{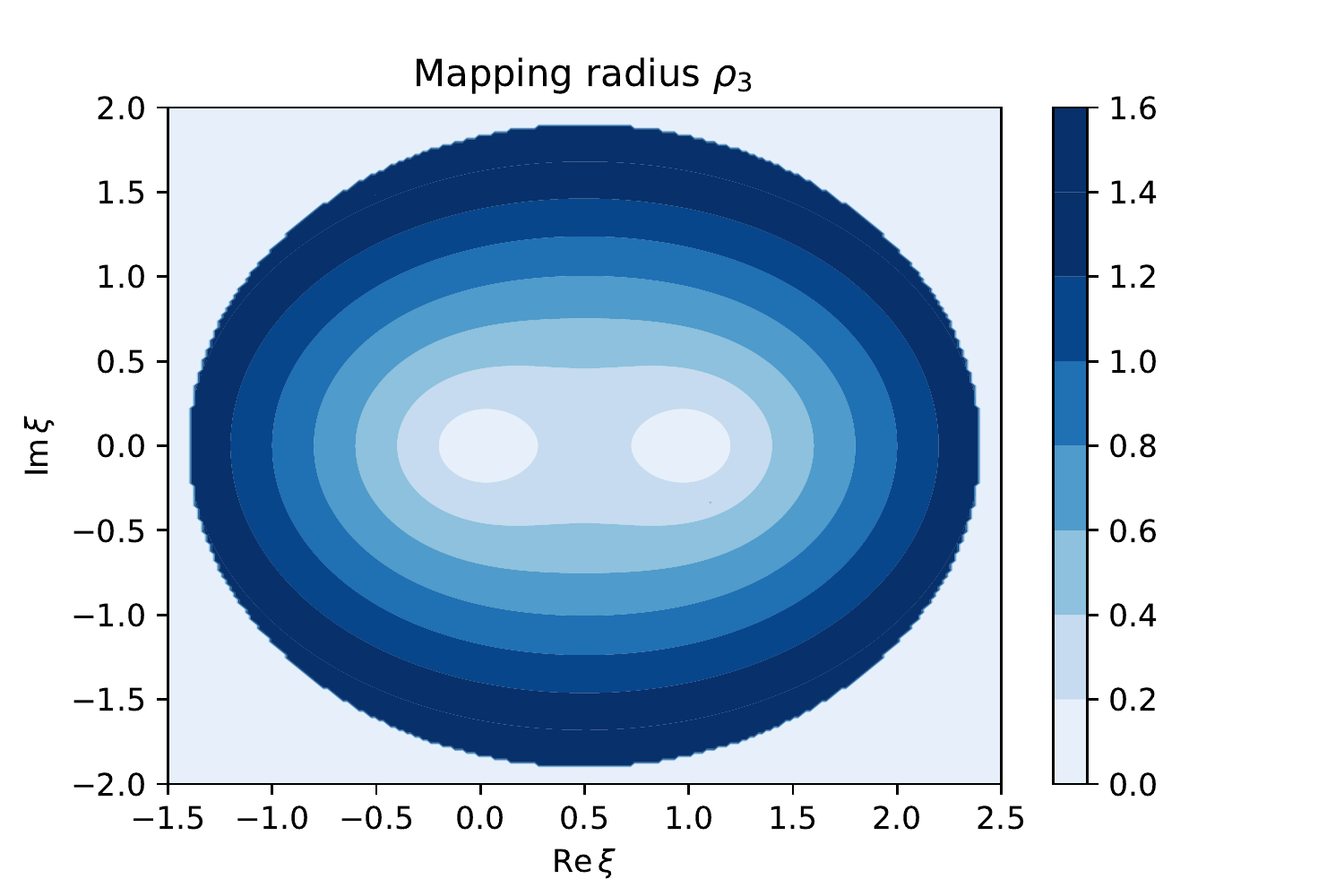}
	\includegraphics[width=8.25cm]{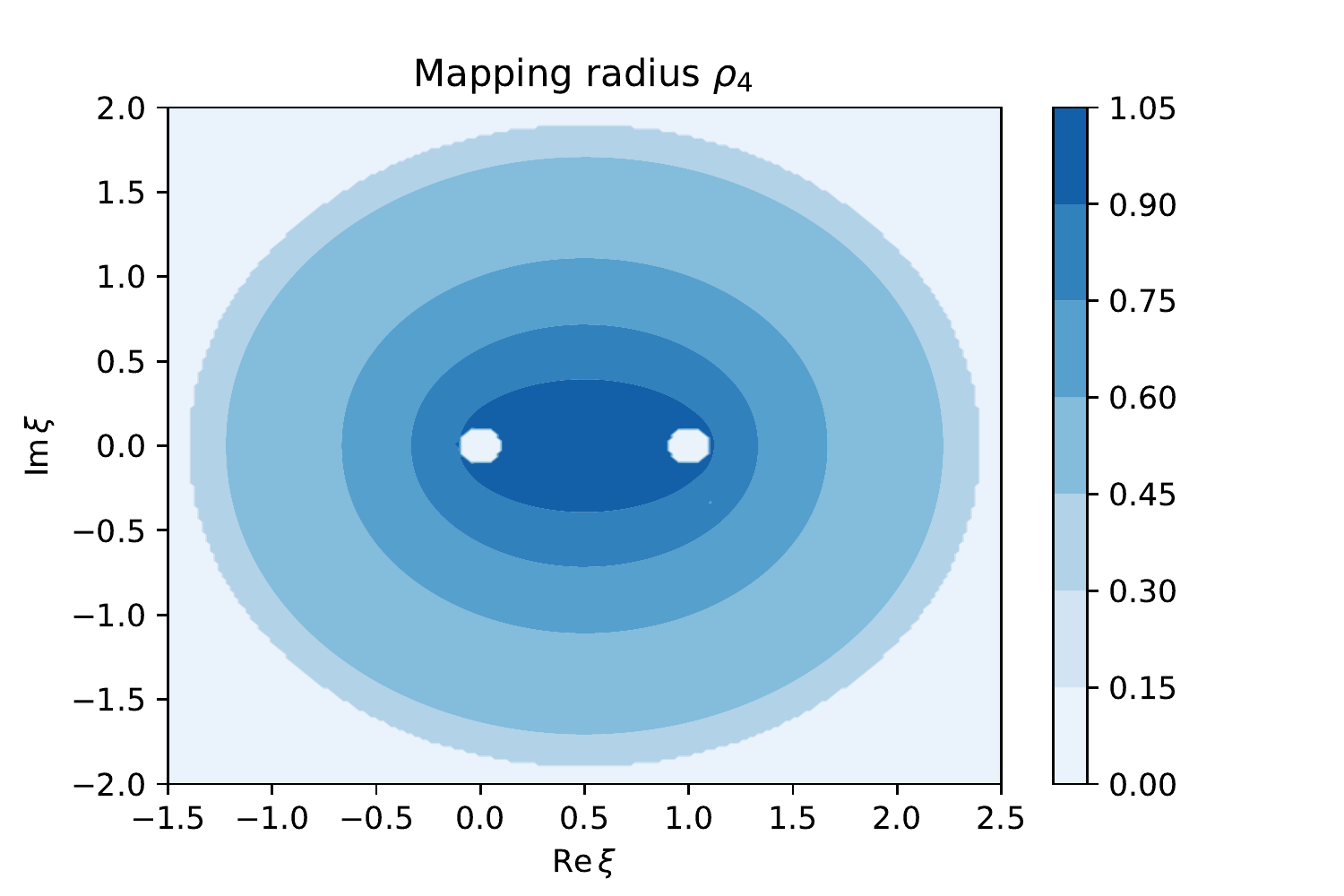}
	\caption{\label{fig:map_rad}The behavior of the mapping radii $\rho_i$ associated with the punctures $\{0,1,\xi,\infty\}$. Small regions around $\{0,1,\infty\}$ excised as the evaluation gets unreliable when punctures are about to collide.}
\end{figure}

The integral~\eqref{eq:AppMap} numerically evaluated using Chebyshev-Gauss method~\cite{abramowitz1964handbook} after changing variables $t = 1-x^2$ using $500$ grid points. Notice that given $\rho_k$, different choices for $z_i$ must give the same result as explained below~\eqref{eq:InvLocCo}. We observed this is indeed the case. For example, the mapping radii and their uncertainties are reported in table~\ref{tab:r} when punctures are placed at $\{0,1,0.8734-0.6242,\infty\}$. The behavior of the (mean) mapping radii is shown in figure~\eqref{fig:map_rad}.
\begin{table}[t]
	\centering
	\begin{tabular}{ | c | c | c | c | c | c | c|}
		\hline
		Map. Rad. & $z_{c,1}$ & $z_{c,2}$ & $z_{c,3}$ & $z_{c,4}$ & Mean & Std. dev. \\
		\hline
		$\rho_1$ & 0.8139  & 0.8126 & 0.8132 & 0.8120 & 0.8129 & 0.0007  \\
		\hline
		$\rho_2$ & 0.4832 & 0.4825 & 0.4821 & 0.4829 & 0.4827  & 0.0004 \\
		\hline
		$\rho_3$ & 0.5176 & 0.5184 & 0.5188 & 0.5180 & 0.5182 & 0.0004  \\
		\hline
		$\rho_4$ & 0.7575 & 0.7569 & 0.7581 & 0.7563& 0.7572 & 0.0007 \\
		\hline
	\end{tabular}
	\caption{\label{tab:r}The mapping radii when punctures are placed at $P = \{0, 1, 0.8734-0.6242i, \infty\}$.}
\end{table}

It was possible to have cancellation errors in the numerical evaluation of the integral~\eqref{eq:AppMap}, but having small standard deviation for the results above indicates that it doesn't actually pose a treat to the accuracy of our computation. We indeed observed that it doesn't cause any issue during our evaluations. In fact, we have used the mean value for the mapping radii in our evaluations and since their associated uncertainties are small we opt to not include them into our final result for $v_4$.

\section{Machine learning details} \label{sec:AppB}

In this appendix, we present more details on our neural networks and their training. We build the training, validation and test sets by randomly sampling $10^5$ points on the complex plane, restricted to the disk of radius $2$ centered at $1/2$ with two disks of radius $0.2$ centered at $0$ and $1$ excised. We use different sets for both networks and each run we sampled new sets. The test sets are left aside until the end of the training to evaluate the performance. The integrals for $v_4$ are computed using yet another set, a grid for trapezoid method and random points in the vertex region for Monte-Carlo.

Both neural networks are fully connected. Training is performed using the eponymous set employing AdamW and we use early stopping: metrics are evaluated on the validation set at the end of each epoch, and training stops when there is no improvement of the loss (resp. accuracy) using the accessory neural network (resp. indicator function neural network) after 100 epochs. Gradients are clipped by the global norm. We employ the following learning rate scheduler: a warm-up period increases the learning rate linearly from 0 to the base learning rate during a given number of epochs; then, the learning rate is decayed exponentially with some period. The hyperparameters are presented in table~\ref{table:hyperparameters} and were found using BOHB hyperparameter optimization~\cite{Falkner:2018:BOHBRobustEfficient}. The statistics for the loss of our runs are shown in table~\ref{table:stats}.
\begin{table}[h]
	\centering
	\begin{tabular}{ | c |  c  |  c |}
		\hline
		Hyperarameter & Accessory parameter NN & Indicator function NN \\
		\hline
		hidden layers & $[512, 128, 1028]$ & $[512, 32, 8, 8]$
		\\
		\hline
		activation function & $\mathbb{C}\text{ELU}(\alpha = 0)$ & ELU($\alpha = 0.25$)
		\\
		\hline
		batch size & 32 & 64
		\\
		\hline
		clip & 6 & 4
		\\
		\hline
		$L_2$ regularization & $8 \times 10^{-6}$ & $4 \times 10^{-2}$
		\\
		\hline
		weight decay & $8 \times 10^{-5}$ & $1.5 \times 10^{-3}$
		\\
		\hline
		base learning rate & $4 \times 10^{-4}$ & $7 \times 10^{-5}$
		\\
		\hline
		warm-up steps & 20 & 5
		\\
		\hline
		decay factor & 0.94 & 0.99
		\\
		\hline
		decay period steps & 5 & 5
		\\
		\hline
	\end{tabular}
	\caption{Hyperparameters of our neural networks.}
	\label{table:hyperparameters}
\end{table}
\vspace{-0.5cm}
\begin{table}[h]
	\centering
	\begin{tabular}{  | c |  l  |  c  |  c  |}
		\hline
		Set & Metric & Mean & Std \\
		\hline
		\multirow{3}{*}{Train}
		& loss (accessory) & $7.11 \times 10^{-8}$ & $6.21 \times 10^{-9}$ \\
		& loss (vertex) & $5.05 \times 10^{-2}$ & $5.89 \times 10^{-3}$ \\
		& accuracy (vertex) & 99.05\% & 0.20\% \\
		\hline
		\multirow{3}{*}{Validation}
		& loss (accessory) & $8.63 \times 10^{-8}$ & $6.19 \times 10^{-9}$ \\
		& loss (vertex) & $5.16 \times 10^{-2}$ & $4.86 \times 10^{-3}$ \\
		& accuracy (vertex) & 98.99\% & 0.18\% \\
		\hline
		\multirow{3}{*}{Test}
		& loss (accessory) & $9.39 \times 10^{-8}$ & $6.12 \times 10^{-9}$ \\
		& loss (vertex) & $3.29 \times 10^{-1}$ & $1.27 \times 10^{-2}$ \\
		& accuracy (vertex) & 99.34\% & 0.18\% \\
		\hline
	\end{tabular}
	\caption{Mean and standard deviations of the metrics for the different data sets, averaged over 4 runs. The full pipeline takes $294 \pm 29$ minutes to run on Google Colab.}
	\label{table:stats}
\end{table}


\providecommand{\href}[2]{#2}\begingroup\raggedright\endgroup

\end{document}